\documentclass[fleqn,usenatbib]{mnras}

\usepackage{newtxtext,newtxmath}

\usepackage[T1]{fontenc}

\DeclareRobustCommand{\VAN}[3]{#2}
\let\VANthebibliography\thebibliography
\def\thebibliography{\DeclareRobustCommand{\VAN}[3]{##3}\VANthebibliography}

\usepackage{graphicx}	
\usepackage{amsmath}	
\usepackage{multicol}

\title[Fornax in $\Lambda{\rm CDM}$]{Can tides explain the low dark matter density in Fornax?}

\author[A. Genina et al.]{
Anna Genina,$^{1,2}$\thanks{E-mail: agenina@mpa-garching.mpg.de (AG)}, Justin I. Read,$^3$ Azadeh Fattahi,$^2$ Carlos S. Frenk$^2$
\\
$^{1}$ Max-Planck-Institut f\"ur Astrophysik, Karl-Schwarzschild-Str. 1, D-85748, Garching, Germany \\
$^{2}$ Institute for Computational Cosmology, Department of Physics, Durham University, South Road, Durham DH1 3LE, UK \\
$^{3}$ Department of Physics, University of Surrey, Guildford, GU2 7XH, UK }

\date{Accepted XXX. Received YYY; in original form ZZZ}

\pubyear{2020}

\begin{document}
\label{firstpage}
\pagerange{\pageref{firstpage}--\pageref{lastpage}}
\maketitle

\begin{abstract}
 The low dark matter density in the Fornax dwarf galaxy is often interpreted as being due to the presence of a constant density `core', but it could also be explained by the effects of Galactic tides. The latter interpretation has been disfavoured because it is apparently inconsistent with the orbital parameters and star formation history of Fornax. We revisit these arguments with the help of the APOSTLE cosmological hydrodynamics simulations. We show that simulated dwarfs with similar properties to Fornax are able to form stars after infall, so that star formation is not necessarily a good tracer of infall time. We also examine the constraints on the pericentre of Fornax and point out that small pericentres (<50 kpc) are not currently ruled out by the data, allowing for Fornax to be tidally influenced on its current orbit. Furthermore, we find that some dwarfs with large orbital pericentres can be stripped prior to infall due to interactions with more massive galaxies. Tidal effects lead to a reduction in the dark matter density, while the profile remains cuspy. Navarro-Frenk-White profiles are consistent with the kinematic data within 3$\sigma$ in the innermost regions, while profiles with shallow cusps or cores provide a better fit. We predict that if the reduction of the dark matter density in Fornax occurs, at least in part, because of the action of Galactic tides, then tidal tails should be visible with a surface brightness limit of $\sim35-36$~mag~arcsec$^{-2}$ over a survey area of $\gtrsim$100~deg$^2$.
\end{abstract}




\section{Introduction}

The $\Lambda{\rm CDM}$ cosmological model has been remarkably
successful in accounting for the cosmic large-scale structure
\citep{fof,percival2001,tegmark2004,millenium}. On small scales,
however, a number of problems persist
\citep{delpopoloreview,weinbergreview, bullockbk}. In particular, the
simplest version of the model predicts that galaxies reside in dark
matter haloes that follow a Navarro-Frenk-White profile (NFW), in
which the density follows a power law ($\rho \propto r^{-1}$) in the
central regions \citep{nfw96,nfw}. This prediction follows from
dark matter-only cosmological $N$-body simulations and has been
recently confirmed over 20 orders of magnitude in halo mass
\citep{wang}. On the other hand, measurements of galaxy rotation
curves in nearby galaxies \citep{Moorecore,
  floresprimack,ohcore,kuzio,adamsrotcurv,oh2015,zhucores,read_curves}
and some analyses of stellar kinematics in nearby dwarf spheroidals
\citep{battagliasculptor,walkerPenarrubia,amoriscoevans,agnelloevans}
appear to favour a `core' in the inner regions, where
$\rho \propto r^0$. This has become known as the `core-cusp problem'
\citep{corecuspreview}.

One particularly interesting galaxy for the core-cusp
problem, is the Fornax dwarf spheroidal. Fornax is one of the
brightest nearby dwarfs and hence large quantities of spectroscopic
and photometric data have been collected for this galaxy
\citep{battagliafornax,walkersculptor,letartefornax,delpinofornax,kirbygradient,kirbymet}. These
data can be used to place a constraint on its dark matter density distribution. A number of studies using stellar kinematics have
favoured a low density core in Fornax \citep{lokas2002,walkerPenarrubia,
  amoriscoFornaxcore,pascalefornax,Jardel_2012,heating,sanders20}. While several works disagree on whether the data are sufficiently constraining to rule out a central density cusp \citep{strigaricore,richardsonfairbairncusp,
  genina,StrigariFrenkWhite}, all studies to date agree that Fornax has a puzzlingly low inner density as compared to expectations in $\Lambda$CDM for a galaxy of Fornax's stellar mass, by a factor of $\sim 3$ \citep{heating}. Further evidence for a low dark matter density in the central regions of Fornax comes from the survival and present-day positions of its globular clusters. A cored dark matter potential in Fornax permits a wide range of initial conditions for Fornax's GCs, whereas a cuspier potential supports a much narrow range, making it statistically less likely in a Bayesian sense   \citep{goerdt,cole_glob_clust,orkney,leung}. However, a cusp cannot be ruled out by such arguments  \citep[e.g.][]{cole_glob_clust,boldrinifornaxcorecusp,meadows}.

The possibility of a dark matter core in Fornax has encouraged
suggestions for alternative dark matter models
\citep{selfintdm,ultralight,ultralight2,sidmcores,correa_sidm}.
However, a simpler explanation within the $\Lambda{\rm CDM}$ framework
is that feedback from supernovae can drive repeated gravitational potential fluctuations that slowly expand dark matter particle orbits, generating a core
\citep{navarroeke,readgilmore,pontzen,alltheway,amoriscoenergycore}. This `dark matter heating' mechanism has been seen in cosmological $N$-body simulations that attempt to model the multiphase interstellar medium \citep{mashchenko08,governato12,nihaocores,firecores,firecores2}. Key to exciting this process is to resolve gravitational potential
fluctuations driven by gas inflows and outflows. This can happen in
simulations where gas is allowed to cool below $T \sim 10^4$\,K and
reach high density before stars form
\citep{pontzen,dutton19}. However, cores can also form in
simulations where gas is not allowed to cool below $T \sim 10^4$\,K,
provided that the density threshold above which gas can turn into
stars is sufficiently high \citep{alejandrocores}. Simulations where
the threshold is low do not generally produce cores
\citep{sawalapuzzles,noproblem, sownak_cores}.

Recent observations, such as in \citet{read_curves,heating}, appear to support the idea that dark matter is `heated up' in at least some dwarf galaxies. Dark matter heating models predict that star formation in dwarfs should be `bursty' \citep{teyssier13}, in agreement with data from dwarf populations\footnote{Note, however, that the converse is not true: burstiness does not necessarily imply that a dark matter core must have formed, as demonstrated by the cuspy profiles of bursty dwarfs in the APOSTLE and AURIGA simulations \citep{sownak_cores}.} \citep{kauffmann14,sparre17}. Stellar orbits should be heated similarly to the dark matter, which is consistent with the high vertical stellar velocity dispersions observed in dwarfs \citep{leaman12,teyssier13,hirtenstein19}. And, these models predict that galaxies whose star formation shut down long ago should be cuspier, which appears to be the case for nearby dwarfs \citep{heating}. This latter point is perhaps the best evidence for dark matter heating to date, however, there is an important potential caveat.

The work of \citet{heating} used a mixture of nearby dwarf irregular galaxies (dIrrs) and quenched Milky Way dwarf spheroidals (dSphs) for their study of dark matter heating, since they required dwarfs with a wide range of star formation histories. While they carefully tested their methodology for both dIrrs and dSphs on a wide array of mock data \citep{read_curves,gravsphere,readdraco,tobeta}, ideally the dIrrs and dSphs should be treated separately since they have distinct systematics (the former have their dark matter distribution inferred from H{\textsc i} gas rotation curves; the latter from stellar kinematics). The dIrrs in their sample all had sufficient star formation to produce a dark matter core and so they cannot be used on their own to test the idea that galaxies with less star formation should be cuspier. By contrast, almost all of the dSphs in their sample had too little star formation for significant core formation to take place. Only one, Fornax, is both a gas-poor Milky Way satellite and had sufficient star formation to produce a dark matter core, making it a particularly interesting test case.

While Fornax holds the promise of being a rosetta stone for testing dark matter heating models, a concern is its proximity to the Milky Way. This raises the spectre that it could have undergone significant tidal stripping or shocking that act to lower its dark matter
density, potentially mimicking the effect of dark matter heating \citep{readtides,penarrubiatides}. \citet{heating} argue that the recent star formation in Fornax \citep{fornaxdeboer}, and its
typically large pericentre inferred from proper motions
\citep{walkerproper,helmiGaia,fritz}, suggest that Fornax is unlikely
to have been subjected to significant tidal effects. Indeed, the
observations of the surface brightness profile of Fornax and $N$-body
models of its orbit have so far disfavoured such interpretations
\citep{walkerfornax,battagliaFornaxtides,wangfornaxtidal}. \citet{coleman1} hypothesise that the shell structures seen in Fornax could be of tidal origin; however, these structures are relatively young compared to the stellar populations in Fornax, favouring instead a recent merger scenario (\citealt{coleman2,coleman3}, but see also \citealt{batefornax}). On the
other hand, comparisons with cosmological simulations favour a tidally
stripped Fornax \citep{noproblem}, where the tails are faint, but detectable with current surveys \citep{wangFornaxTails,wangfornaxtidal}. Simulations in CDM and binding energy arguments also favour an infall time for Fornax of
$\sim 8$~Gyr ago \citep{rocha,wangfornaxinfall,fillingham_infall},
which suggests that Fornax has spent a long time under the influence
of tides. Finally, even if Fornax's apparent orbit today suggests relative tidal isolation, its orbit and tidal field may have been rather different in the past, especially if it was pre-processed inside an infalling group \citep{lux10}.

In this paper, we set out to test whether the inferred dark matter density in
Fornax can be reproduced in a $\Lambda{\rm CDM}$ cosmological $N$-body hydrodynamics simulation in which dwarf galaxies do not form cores. In Section~\ref{sec2}, we briefly describe our suite of
simulations. In Section~\ref{sec3}, we identify appropriate analogues
of Fornax given its dynamical mass, luminosity, star-formation
history and the most recent orbital constraints from {\it Gaia}. In
Section~\ref{sec4}, we analyse the formation pathways of the Fornax
analogues and compare their inner densities to that inferred for
Fornax. In Section~\ref{sec5}, we outline the possible limitations of our model for the tidally-induced reduction in the central density of Fornax. In Section~\ref{sec6}, we discuss the significance of our results for the core-cusp problem in Fornax. Finally, in Section~\ref{sec7} we present our conclusions.

\section{Simulations}
\label{sec2}

We use the APOSTLE suite of simulations of Local Group-like
environments \citep{fattahi,sawalapuzzles}. APOSTLE is based on the
{\sc eagle} galaxy formation model and includes models for hydrogen and helium reionization, star formation, gas cooling, metal enrichment, supernovae feedback, AGB winds and AGN (see \citealt{eaglecrain} and \citealt{eagle} for full details). {\sc eagle} is an extension of the {\sc p-gadget-3} code, which itself is an improvement on the publicly available {\sc gadget-2} \citep{gadget}. The hydrodynamics are implemented using the
pressure-entropy formalism of \citet{hopkins}. In APOSTLE, the
`reference' {\sc eagle} model is used. APOSTLE assumes a
$\Lambda {\rm CDM}$ model in the WMAP-7 cosmology \citep{wmap7}. The suite
consists of three resolution levels and, in this work, we use the five
highest resolution volumes, with dark matter particle masses ranging
between m$_{\rm{DM}} = 2.5-5\times10^4$M$_{\odot}$ and gravitational 
softening parameter, $\epsilon=134$~pc. Gas particles have initial masses in
the range $5-10\times10^3$M$_{\odot}$.  

Each APOSTLE volume includes two main haloes, which match the Milky Way and Andromeda pair in terms of their total mass, physical separation, relative velocities and the surrounding Hubble flow \citep{fattahi}. Each of the main
haloes hosts a population of dwarf galaxies. The volumes also include nearby isolated dwarfs. Dwarfs in APOSTLE have been extensively
studied in terms of their morphological and dynamical properties
\citep{campbell}, star formation histories \citep{digby}, metallicities and metallicity gradients (including distinct metallicity subpopulations), and gas content \citep{thedistinct}. These properties were shown to be consistent, within the resolution limits, with Local Group dwarfs. 

In this work, in order to identify bound substructures, we employ the
{\sc hbt+} algorithm of \citet{hbt}. {\sc hbt+} makes it particularly
straightforward to track haloes across time by assigning each halo a unique `TrackID'.

\section{Methods}

\begin{figure}

		\includegraphics[width=\columnwidth]{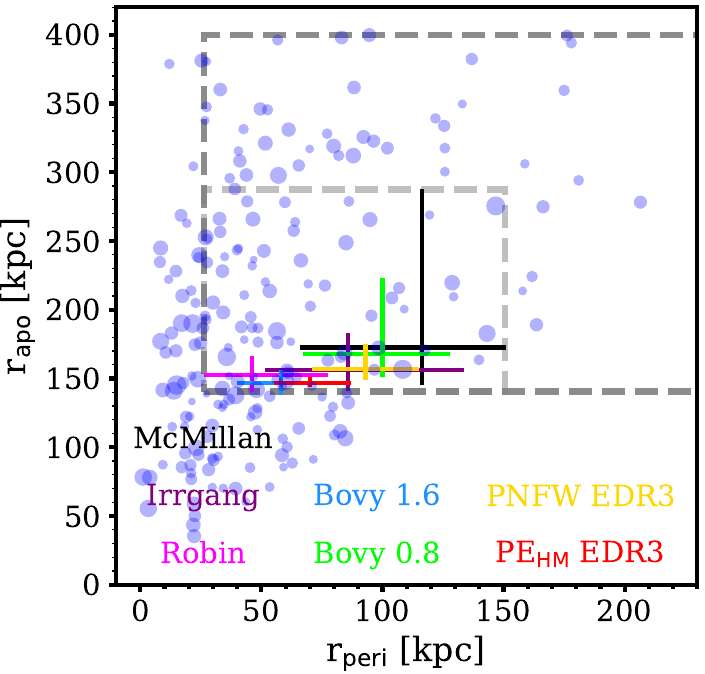}
    	
		\caption{The inferred apocentres and pericentres for
                  all dwarfs within the APOSTLE volumes, which are
                  currently located within 300~kpc of the centre of
                  Milky Way analogues. The area of each point is
                  proportional to the logarithm of the stellar mass at $z=0$.
                  The labelled error bars mark constraints for Fornax from various
                  potentials described in \citet{helmiGaia},
                  \citet{fritz} and \citet{gaia_edr3_pm}. The light grey box shows the
                  selection of our sample of Fornax-like orbits and
                  the dark grey box shows a wider selection of dwarfs
                  that have even larger apocentres than those
                  predicted for Fornax.}

		\label{fig1}
	\end{figure}

\label{sec3}

Our aim is to investigate mass loss in simulated dwarfs
similar to Fornax. The simulated Fornax analogues are required to match
Fornax according to the following criteria: {\it i)} an orbit that is
compatible with the {\it Gaia} data; {\it ii)} halo mass and
luminosity; {\it iii)} star formation history. In the
following, we outline the methods behind our sample selection.

\subsection{Sample selection}

As a first step, we select satellite dwarfs from our simulations,
which we define as galaxies that are presently within 300~kpc of
either the Milky Way or Andromeda analogues in APOSTLE. Of those, we identify the objects with a stellar mass of at least $10^5 M_{\odot}$, corresponding to stellar populations resolved with $\sim 100$ particles. We further check that the tidal evolution of these subhaloes is well resolved. Thereafter, we keep all dwarfs which satisfy the criteria in eqn.~16 of \citet{vandenbosch2} and have at least 1000 bound dark matter particles at $z=0$. While the tidal evolution of subhaloes down to $\sim$300~particles can, in principle, be resolved according to their eqn.~19, this would still correspond to 0.1~dex scatter in the fractions of bound particles surviving in each of the possible realizations of a dark matter halo. The 1000-particle resolution limit corresponds to dark matter halo masses above a few times $10^7$~M$_{\odot}$ and leaves us with a sample of 212 resolved luminous dwarfs. As we will show in Section~\ref{heating_comp}, the pre-infall halo masses of our sample of Fornax analogues are in the range $4\times10^9 - 1\times 10^{10} M_{\rm \odot}$, corresponding to $\gtrsim10^5$ dark matter particles at infall. These simulated analogues lose no more than 90~per~cent of their maximum mass and are thus robust to the effects of numerical disruption at present day, even for orbital pericentres as small as 10~per~cent of the virial radius of the host halo (see fig.10 of \citealt{vandenbosch2}, where the yellow band highlights the gravitational softening choice of \citealt{power}, applicable to our simulations).

\subsection{Orbits in the host potential}
Next, we
approximate the potential of each host galaxy as being spherically
symmetric. The potential for a spherically symmetric mass
distribution is given by \citep{formationevolution}: 
\begin{equation}
    \Phi(r) = -G\int_r^{\infty} dr' \frac{M(<r')}{r'^{2}}.
\end{equation}
We compute the potential out to a 
radius of 400~kpc, where we fix $\Phi(400 \rm ~kpc) = 0$. We thus only consider simulated dwarf galaxies with the maximum apocentre of 400~kpc.

For each dwarf we compute the total energy, $E$, and the angular
momentum, $L$. Then, for the orbital extremes, where the velocity
$\dot{r} = 0$, we have that:

\begin{equation}
    \dot{r}^2 = 2[E-\Phi(r)] - \frac{L^2}{r^2} = 0.
    \label{eqperiapo}
\end{equation}

The solutions of this equation correspond to the apocenter,
$r_{\rm apo}$, and the pericenter, $r_{\rm peri}$, of the orbit
\citep{galacticdynamics}. We verify that the approximation of the
potential as spherically symmetric is valid in Appendix~A.
 
In Fig.~\ref{fig1}, we show the distribution of orbital pericentres and
apocentres for all satellite dwarfs with $M_* > 10^5$~M$_{\odot}$. The
error bars of various colours represent constraints for Fornax using three Milky Way models from \citealt{helmiGaia} (using {\it Gaia DR2}), two Milky Way
models from \citealt{fritz} (using {\it Gaia DR2}) and two models from \citet{gaia_edr3_pm} (using {\it Gaia EDR3}). The first three are the potentials from
\citealt{McMillan} (M$_{200}=1.3\times10^{12}$ M$_{\odot}$), \citealt{Irrgang2012:1211.4353v5} (M$_\mathrm{< 200~ kpc}$ = $1.9\times10^{12}$ M$_{\odot}$) and
\citealt{robin2003, robin2012} (M$_\mathrm{< 100~ kpc}$ = $1.2 \times 10^{12}$ M$_{\odot}$). The three potentials differ most
significantly in the inner $\sim2$~kpc and beyond $\sim40$~kpc. {\it
  Bovy 1.6} and {\it Bovy 0.8} are two realizations of the potential
from \citet{bovy} with virial masses, $M_{200}$, of
$1.6\times10^{12}M_{\odot}$ and $0.8\times10^{12}M_{\odot}$,
respectively. Finally, we include pericentre and apocentre constraints using {\it Gaia EDR3} from \citet{gaia_edr3_pm}, where a "high-mass" potential from \citealt{retana-montenegro} (PE$_{\rm HM}$) and a "low-mass" potential from \citealt{pouliasis} (PNFW) were used, which are roughly similar in mass to the two models from \citet{fritz}. In our sample, we include all orbits contained within the
error bars of these models. This is shown as a light grey box in
Fig.~\ref{fig1}. Additionally, we include dwarfs with larger peri- and
apocentres, as we wish to investigate mass loss for these seemingly
`benign' orbits. This additional selection is highlighted with a dark
grey box in Fig.~\ref{fig1}.

\subsection{$V_{1/2}$ - $M_V$ relation}

	\begin{figure*}
   
    \begin{multicols}{2}
         \includegraphics[width=\columnwidth]{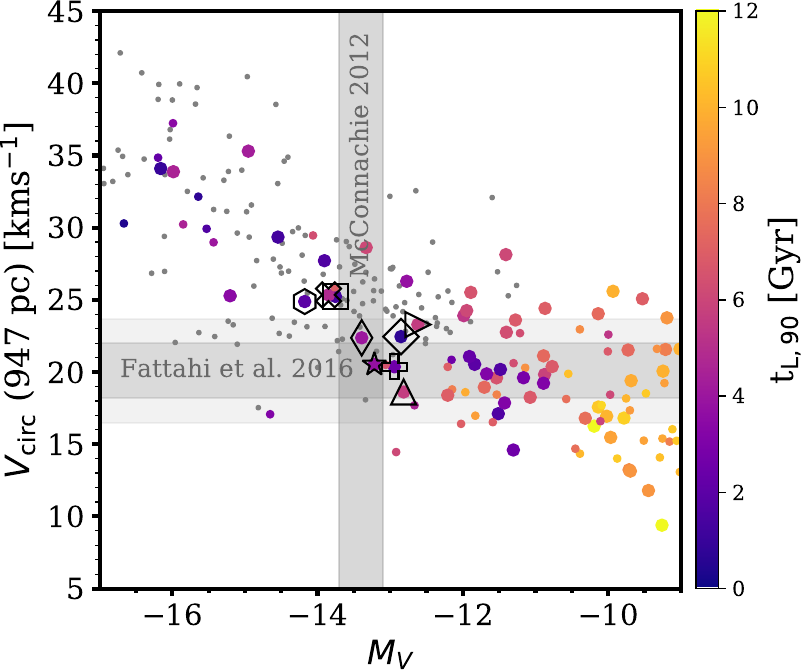}
        
		\includegraphics[width=.9\columnwidth]{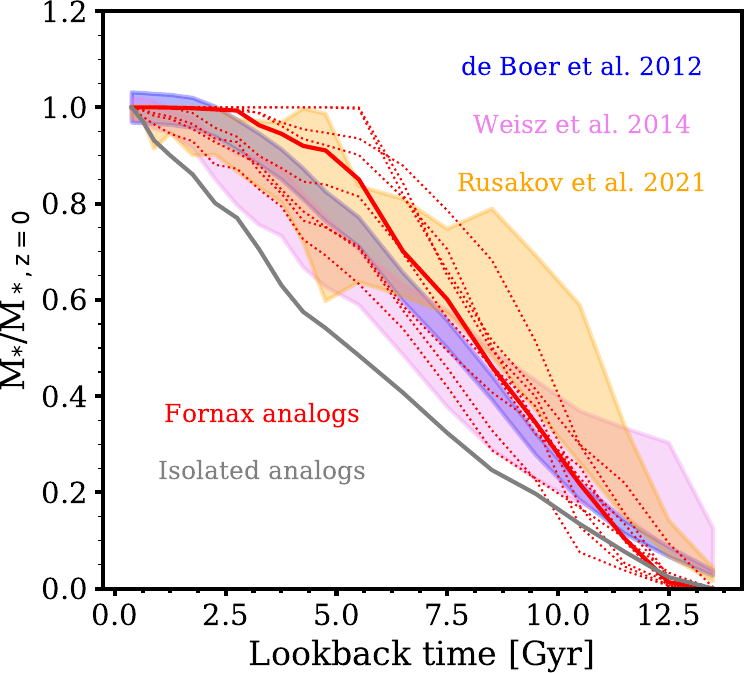}
    \end{multicols}    
    \caption{{\it Left:} the circular velocity at the half-light radius of Fornax, $V_{\rm circ}(\rm{947~pc})$, as a function of the absolute visible
      magnitude, $M_V$. The dark grey and light grey horizontal bands span the 1$\sigma$ and 2$\sigma$ errors on the value of $V_{1/2}$ for Fornax from \citealt{noproblem,fattahi_tides} ($V_{1/2}=20.1^{+1.9^{+3.6}}_{-1.9_{-3.6}}$ kms$^{-1}$), while the vertical grey band spans the $1\sigma$ error on the absolute visible magnitude ($M_V=-13.4\pm0.3$) from \citet{mcconnachieCensus}. The thick circles are dwarfs with orbits
      comparable to Fornax and the thin dots are other satellite
      dwarfs. Grey dots are isolated dwarf galaxies. The sample of
      satellites is coloured by the lookback time at which they formed
      90~per~cent of their $z=0$ stellar mass, $t_{\rm L,90}$. Symbols of various shapes mark our
      9 best Fornax analogues. {\it Right:} the normalized star
      formation history of Fornax from
      \citealt{fornaxdeboer} (blue band), \citealt{weisz} (pink band) and \citealt{rusakov} (orange band, where we used the upper an lower bands of both Fornax models). The thin red lines show histories for our
      Fornax analogues and the thick red line is their median. The
      thick grey line is the median for isolated dwarfs which are similar to Fornax in terms of $V_{\rm circ}(\rm{947~pc})$ and $M_V$.}

		\label{fig2}
	\end{figure*}
	
Now that the satellite galaxies with orbits similar to Fornax have
been identified, we wish to select those which resemble
Fornax in their present-day stellar and halo masses. Since neither of
these two properties can be measured directly, we use the circular
velocity at the deprojected half-light radius of Fornax, $V_{1/2} \approx V_{\rm circ} (\rm{947~pc})$, and the absolute
magnitude in visible light, $M_V$, as indicative quantities. $V_{\rm circ}(\rm{947~pc})$
is computed as $ \sqrt{\frac{GM(<\rm{947~pc})}{\rm{947~pc}}}$ , where
947~pc is $4/3$ times the half-light radius of Fornax (710~pc), which is approximately the 3D half-light radius \citep{wolf} and $G$ the gravitational
constant. We obtain the
visible magnitudes for our dwarfs using the {\sc galaxev} stellar population synthesis model \citep{galaxev} applied to all bound stellar particles, as determined by the {\sc subfind} algorithm. Our method is described in more detail in \citet{thedistinct}. For comparison with Fornax, we use the
corresponding quantities for $V_{1/2}=20.1^{+1.9^{+3.6}}_{-1.9_{-3.6}}$ km~s$^{-1}$ from \citealt{noproblem} (where inner errors denote the 25-75$^{\rm th}$ percentiles of the distribution, and the outer errors the 10-90$^{\rm th}$ percentiles) and $M_V=-13.4\pm0.3$ from
\citet{mcconnachieCensus}. 

The values of $V_{\rm circ}(\rm{947~pc})$ and $M_V$ for our sample of dwarfs are shown in the left panel of Fig.~\ref{fig2}, where large colour circles
represent the galaxies contained within the dark grey box in
Fig.~\ref{fig1}, small circles are other satellite galaxies and grey
dots are isolated dwarf galaxies. The satellite dwarfs are coloured by
$t_{\rm L,90}$, the lookback time at which 90~per~cent of their $z=0$ stellar particles were formed. It can be seen that above the $M_V$ of Fornax, the
majority of satellites are still star-forming today. 

It is clear from the left panel of Fig.~\ref{fig2} that Fornax is,
indeed, somewhat of an outlier in the APOSTLE $V_{\rm circ}(\rm{947~pc})$ - $M_V$ relation. Nevertheless, we select a sample of 9 similar galaxies for
further inspection. These are marked with different shapes in
Fig.~\ref{fig2}, which we will continue using in the remainder of this
paper. It can be seen that the `star' dwarf is our best Fornax
analogue. There is a larger number of isolated dwarfs that match these
constraints for Fornax; however, as we shall see below, their star
formation histories are incompatible with that of Fornax.

\subsection{Star formation histories}

We would now like to ensure that our simulated analogues are indeed
similar to Fornax in terms of their star formation history (SFH). For our simulated dwarfs, we bin the formation times of all bound stellar
particles into the same set of bins as in \citet{fornaxdeboer}. The right panel of Fig.~\ref{fig2} shows the cumulative SFHs of these galaxies as a
function of lookback time, normalized by the present-day stellar
mass. The solid red line is the median of the 9 simulated Fornax
analogues. The blue shaded band is the observational result from
\citet{fornaxdeboer}, pink from \citet{weisz} and orange from \citet{rusakov}.

Our Fornax analogues, on average, exhibit a similar shape in their SFHs
as Fornax. There are two examples that have stopped star formation
$\sim6$~Gyr ago. This is clearly at odds with Fornax, which, at most,
stopped making stars $\sim2$~Gyr ago. In the next Section, we will
investigate whether this ``quenching time'' is indicative of the tidal
history of this dwarf in our simulations.

The thick grey line on the right panel of Fig.~\ref{fig2} shows the median
SFHs of isolated dwarfs. It is
clear that these galaxies are still star-forming today and many have an
increased star formation rate over the past 5~Gyr. This is
incompatible with Fornax whose star formation rate is slowing down
around that time. The SFH of Fornax is certainly more compatible with the simulated dwarfs currently within the virial radius of the Milky Way analogue than with
isolated dwarfs.

Now that our sample of dwarf galaxies has been selected we proceed
with our analysis and results.

\section{Results}
\label{sec4}

\begin{figure*}
        \begin{multicols}{2}
		\includegraphics[width=\columnwidth]{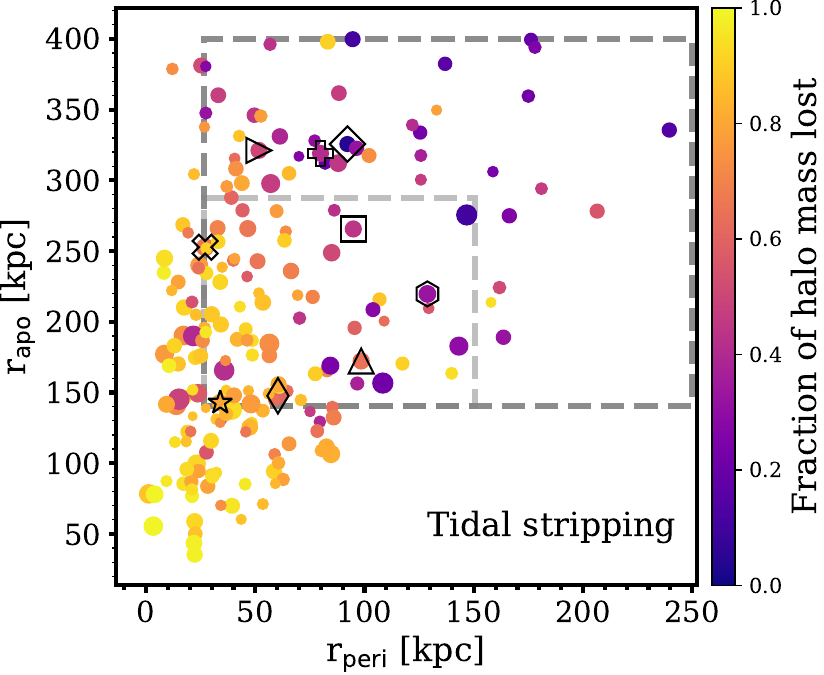}\par
		\includegraphics[width=\columnwidth]{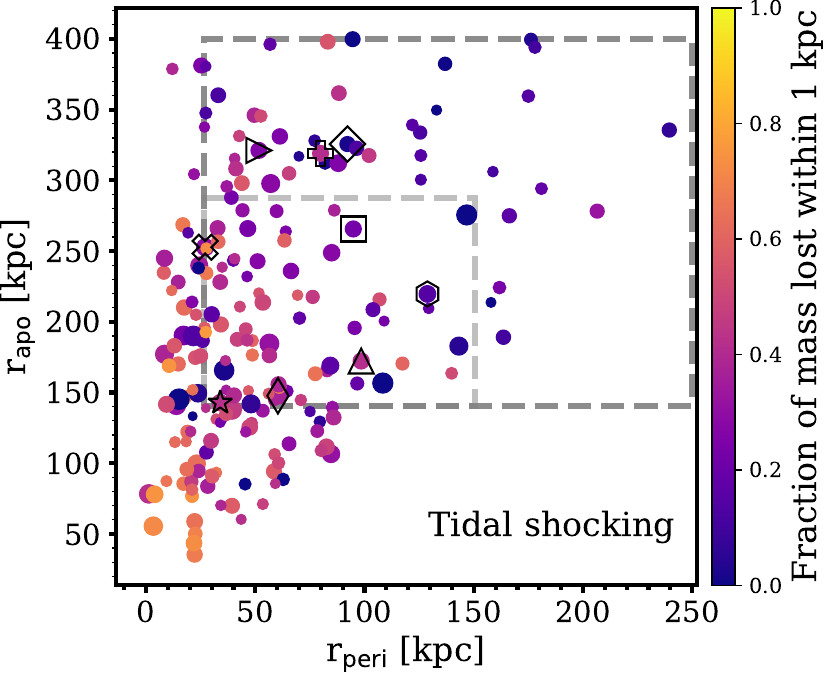}\par
	    \end{multicols}

            \caption{{\it Left:} the inferred apo- and pericentres of
              simulated dwarfs, coloured according to the fraction of
              peak halo mass lost,
              indicative of tidal stripping. The area of each circle
              is proportional to the logarithm of the stellar mass at $z=0$.
              The grey dashed boxes are as in Fig.~\ref{fig1}. The
              symbols with shapes mark our 9 best Fornax
              analogues. {\it Right:} as on the left, but coloured
              according to the mass lost by each dwarf within the 
              central 1~kpc, indicative of tidal shocks.}
    \label{fig3}
		
	\end{figure*}
\begin{figure*}
        \begin{multicols}{2}
		\includegraphics[width=\columnwidth]{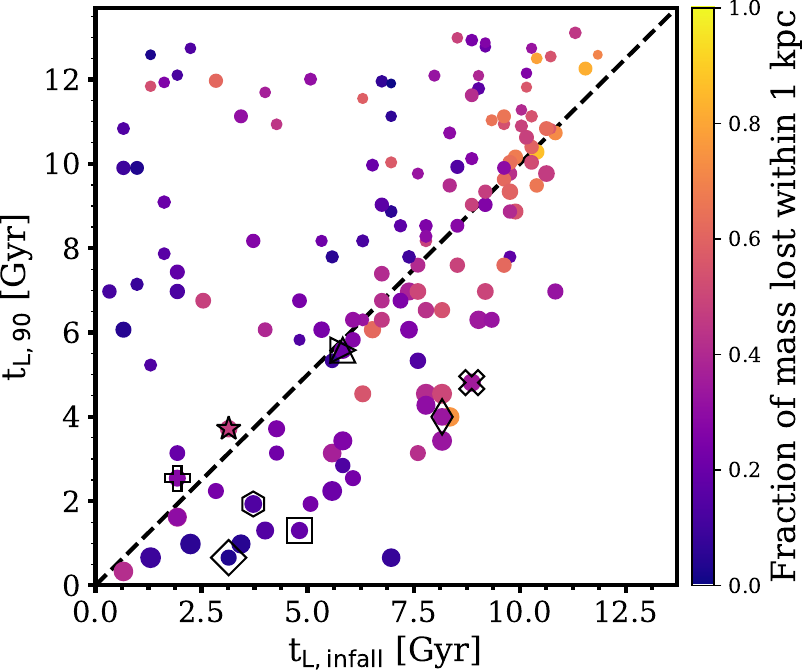}\par
		\includegraphics[width=\columnwidth]{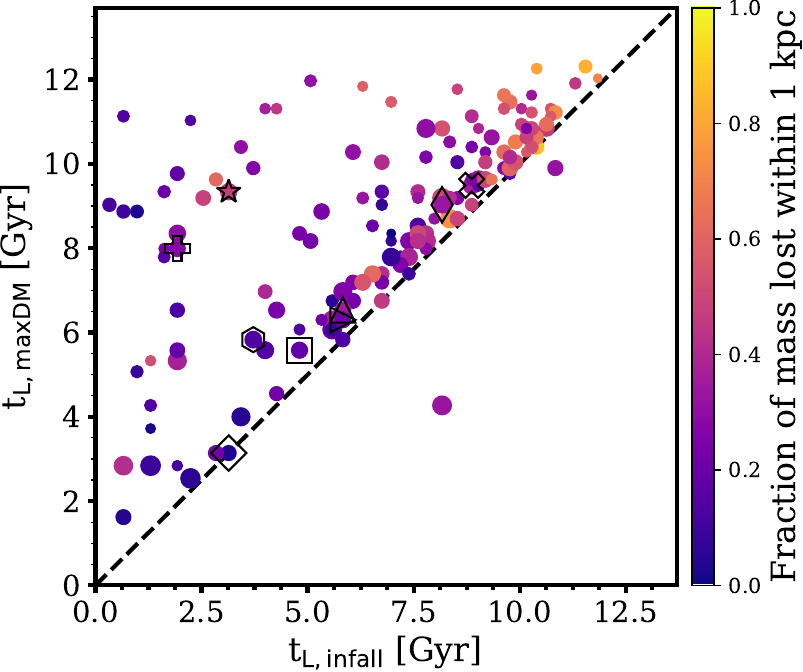}\par
	    \end{multicols}

            \caption{{\it Left:} the lookback time at which each
              galaxy forms 90~per~cent of its stars, $t_{\rm L,90}$, as a
              function of the lookback time when it first crosses the
              virial radius of the host halo, $t_{\rm L,infall}$. The
              points are coloured by the fraction of mass lost within
              1~kpc. The area of the circles represents the logarithm
              of the $z=0$ stellar mass. {\it Right:} the lookback time at
              which each galaxy had its maximum dark matter halo mass
              as a function of the infall time, $t_{\rm
                l,infall}$. The points are coloured by the mass lost
              within 1~kpc. In both plots the symbols mark the 9 best
              Fornax analogues and the black dashed lines show the one-to-one relation.}
    \label{fig4}
		
	\end{figure*}

\subsection{Does the current orbital pericentre reflect tidal history?}

It has been argued by \citet{heating} that Fornax is unlikely to have
undergone significant tidal stripping because it has a relatively
large pericentre and is on a nearly circular orbit. These inferred
properties, of course, depend on the potential chosen to model the
Milky Way. Given the various pericentre values in different potentials
from \citet{helmiGaia}, \citet{fritz} and \citet{gaia_edr3_pm}, we examine whether this
argument holds in our simulations. We note that the stellar mass, mostly comprised of the disk and bulge, of
the Milky Way and Andromeda analogues in APOSTLE is {\it smaller} by a factor of $\sim$2.5-5 than
the stellar mass inferred for the Milky Way \citep{fattahi, shaken}, while the virial
masses of the haloes, $M_{200}$, are in the range of values for the Milky Way quoted in the literature \citep{wang_review}. It is believed that satellite disruption can be enhanced due to interactions with Milky Way's stellar disk \citep{onghiadisk}. This implies that, for a given orbit, on average, the tidal effects are expected to be {\it less severe} in our simulated potentials than in the Milky Way.

We track each of the APOSTLE satellites back to the time when their
dark matter halo mass reached its peak value. In Fig.~\ref{fig3}, we show the
fraction of bound dark matter mass that has been lost since that time according to the {\sc subfind} algorithm
(left), indicative of tidal stripping, and the dark matter mass lost
from within 1~kpc from the centre of each dwarf, indicative of tidal
shocks. We note that the mass lost within 1~kpc does not necessarily reflect mass leaving the halo, although this quantity does reflect the reduction in the central density\footnote{We have verified that a fraction of particles present within 1~kpc at infall is removed completely from the halo, though this is typically a smaller fraction than that displayed on the right of Fig.~\ref{fig3}.}. It can be seen that the vast majority of dwarfs with
pericentres below 50~kpc lose more than 80~per~cent of their halo
mass. Even dwarfs that come only as close as 150~kpc from the centre of
the Milky Way lose significant fractions of their mass. Galaxies with
large pericentres (> 150 kpc) and less eccentric orbits lose very
little mass. 

Mass loss from the innermost 1~kpc appears to require more extreme
conditions. Galaxies with pericentres <50~kpc and apocentres <100 kpc
tend to lose more than 80~per~cent of their halo mass within 1~kpc. At
larger peri- and apocentres the dwarfs typically lose no more than
50~per~cent of their mass within 1~kpc.  There are, however, some
clear exceptions to the trend. In particular, it is interesting that a
number of dwarfs with pericentres as large as 100~kpc can still lose
large fractions of their innermost dark matter mass.

The shape symbols, as in Fig.~\ref{fig2}, show our Fornax
analogues. We can see that all of these have pericentres below
100~kpc. Let us look at, for example, the `upwards triangle' Fornax. This
galaxy is on a fairly non-eccentric orbit, with a pericentre of
$\sim100$~kpc, and matches the constraints for Fornax rather
well. Nevertheless, it loses $\sim 70$~per~cent of its halo mass and
$\sim 40$~per~cent of its mass within 1~kpc. Clearly, even for
non-eccentric, large-pericentre orbits, the galaxies can lose large
fractions of dark matter and this phenomenon is not infrequent. Within
the light grey box, on the right panel of Fig.~\ref{fig3}, 26~per~cent
of dwarfs lose more than 50~per~cent of their dark matter mass within
1~kpc. We will return to the implications of this for the dark matter
density of Fornax in Section~\ref{heating_comp}.

\subsection{Does recent quenching indicate recent infall?}

We now consider whether the quenching time is a good indicator of the
infall time of a dwarf galaxy and, consequently, an indicator of how
much the galaxy has been affected by tides.  

On the left panel of Fig.~\ref{fig4} we show the quenching time of
each dwarf galaxy , $t_{\rm L,90}$, as a function of its infall time,
$t_{\rm L,infall}$. We define the infall time as the time when the
galaxy first crosses the virial radius of a simulated Milky Way analogue. It can be seen that a fraction of simulated dwarfs closely trace a one-to-one relation (black dashed line) between the infall and quenching times (i.e. many satellites are quenched immediately after
infall). However, it can also be seen that a number of dwarf galaxies
are quenched prior to infall and a number of them after infall. 

The area of the points in Fig.~\ref{fig4} is proportional to the
logarithm of the present-day stellar mass. It is thus clear that there
exists a population of low-mass dwarfs that are quenched prior to
infall because of reionization and feedback from star formation \citep[see also][]{digby}. In fact, it can be seen that the smallest galaxies have very early quenching times ($\sim12$~Gyr) and, as the mass increases, for a given
$t_{\rm L,infall}$, quenching is delayed. On the other hand, there
exists a population of dwarf galaxies that are generally more massive
(closer to Fornax in mass), where quenching occurs after infall
(because of ram pressure stripping and tidal stripping of gas inside
the host halo). Moreover, as discussed by \citet{thedistinct}, a
number of dwarfs with large gas supplies are able to form more stars
near the pericentre of their orbits\footnote{We point out that the distance--$\rm HI$ relation for APOSTLE dwarfs is in agreement with Local Group dwarfs (see the Appendix of \citealt{thedistinct}). Additionally, \citealt{kelly_apostle} and \citealt{kelly_eagle} find that the hot halo masses of main galaxies in APOSTLE are compatible with those inferred for the Milky Way \citep{millerbregman}.}. If gas remains after the first
pericentric passage, another star formation burst may occur near the next
pericentre. As such, it is clear that, for higher mass objects, the
quenching time is not a good indicator of the infall time. Moreover,
from the colour of the points (representing the fraction of mass lost
within 1~kpc), it can be seen that a number of satellites that are
quenched after infall also lose large fractions of their dark matter
mass.

On the right of Fig.~\ref{fig4} we show the time at which the galaxy
had its maximum dark matter mass, $t_{\rm L,maxDM}$, as a function of
infall time, $t_{\rm L,infall}$. In principle, these quantities should
have a very similar value. Indeed, we see that for a large number of
dwarfs the relation is nearly one-to-one (black dashed line). However, there also exists a
population of galaxies that start losing mass long before
infall. Among these objects are some of our best Fornax analogues,
including the `star' Fornax, which is a match in both $V_{\rm circ}{(\rm{947~pc})}$ and $M_V$. We also note the presence of the `plus` Fornax in this region, however, as we shall see later, this galaxy simply stops halo assembly $\approx 8$~Gyr ago and only experiences minor mass loss through tidal interactions prior to infall. We find that among the population of satellites with orbits compatible with Fornax (light grey box in Fig.~\ref{fig3}), nearly 37~per~cent have started losing mass more than 2~Gyr prior to infall. From the colours of these galaxies, indicating mass loss
within 1~kpc, it is clear that these objects can lose up
to $\sim 90-99$~per~cent of their maximum halo mass, despite very
recent infall times in some cases. We conclude that the infall time is
not always an adequate indicator of a dwarf's tidal history.

\subsection{The diverse formation histories of Fornax analogues}

\begin{figure*}
    \centering
    \begin{multicols}{2}
     \includegraphics[width=\columnwidth]{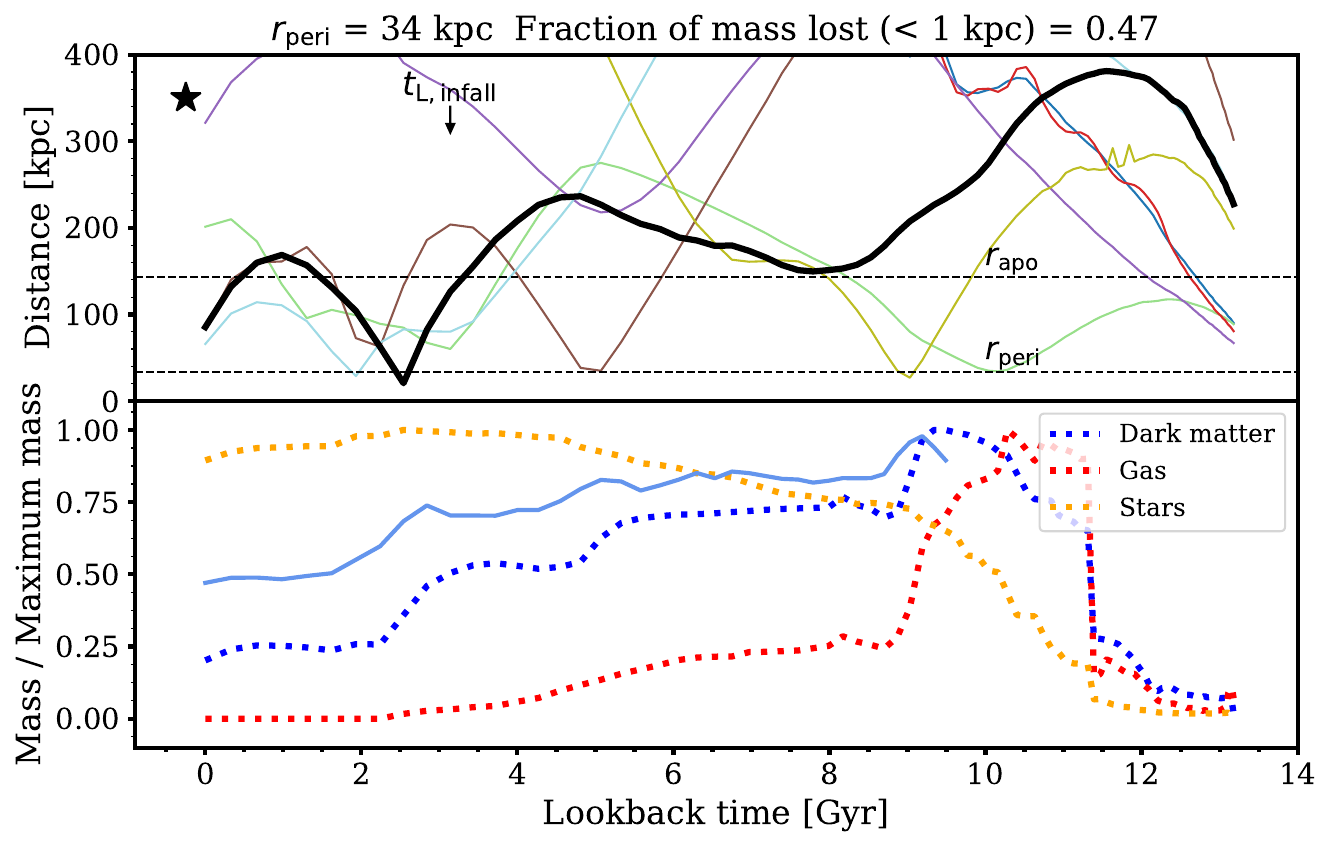}\par
    \includegraphics[width=\columnwidth]{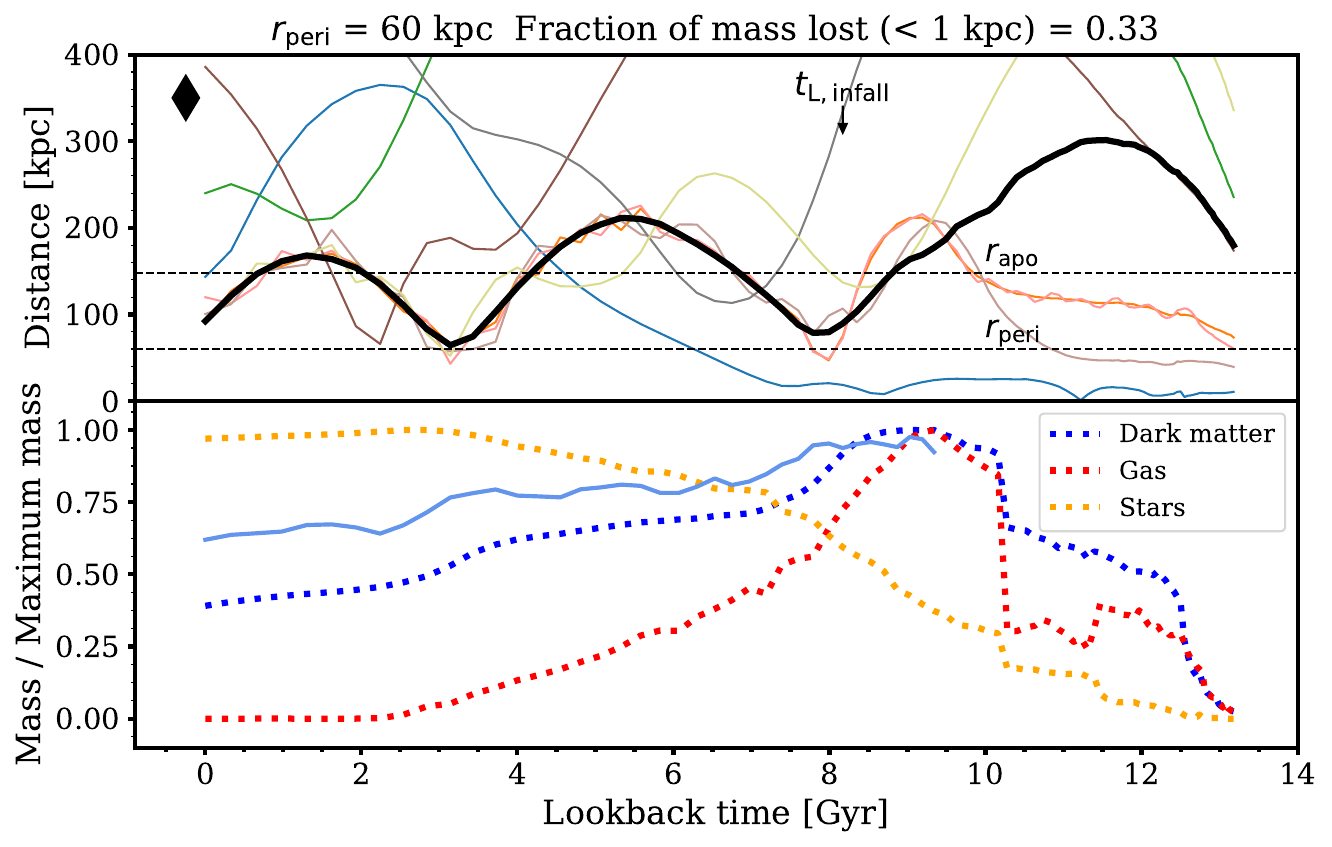}\par

	    \includegraphics[width=\columnwidth]{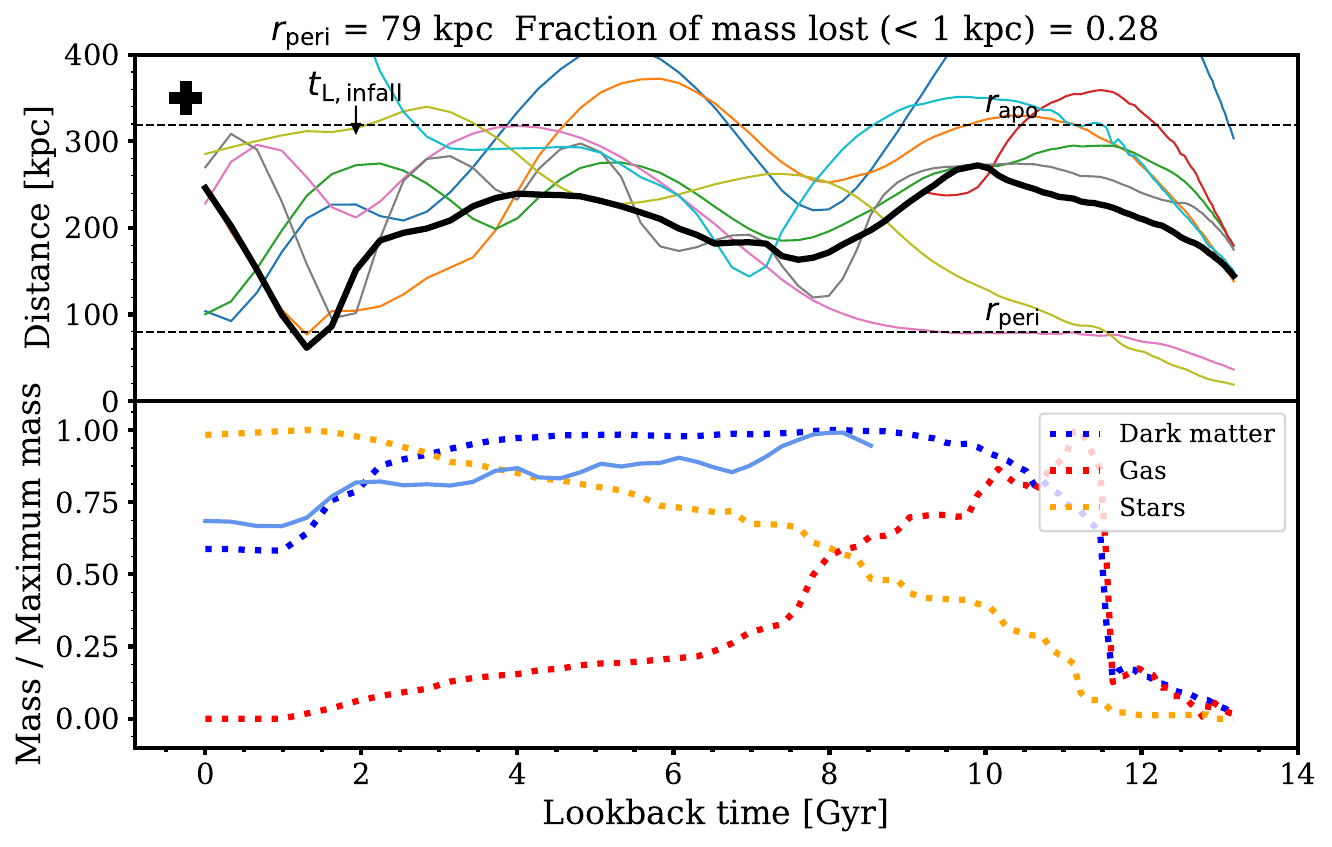}\par 
 
      \includegraphics[width=\columnwidth]{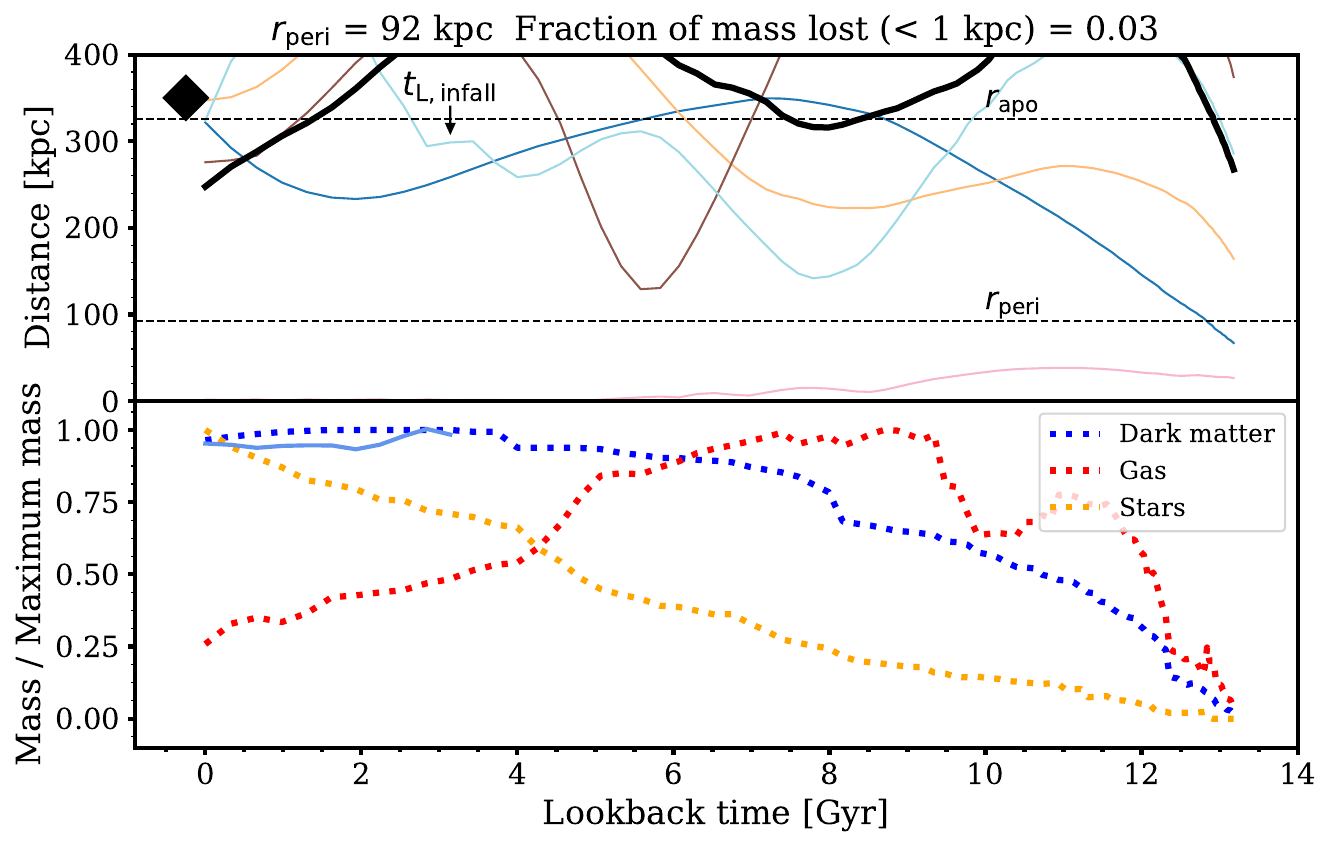}\par
    \includegraphics[width=1.\columnwidth]{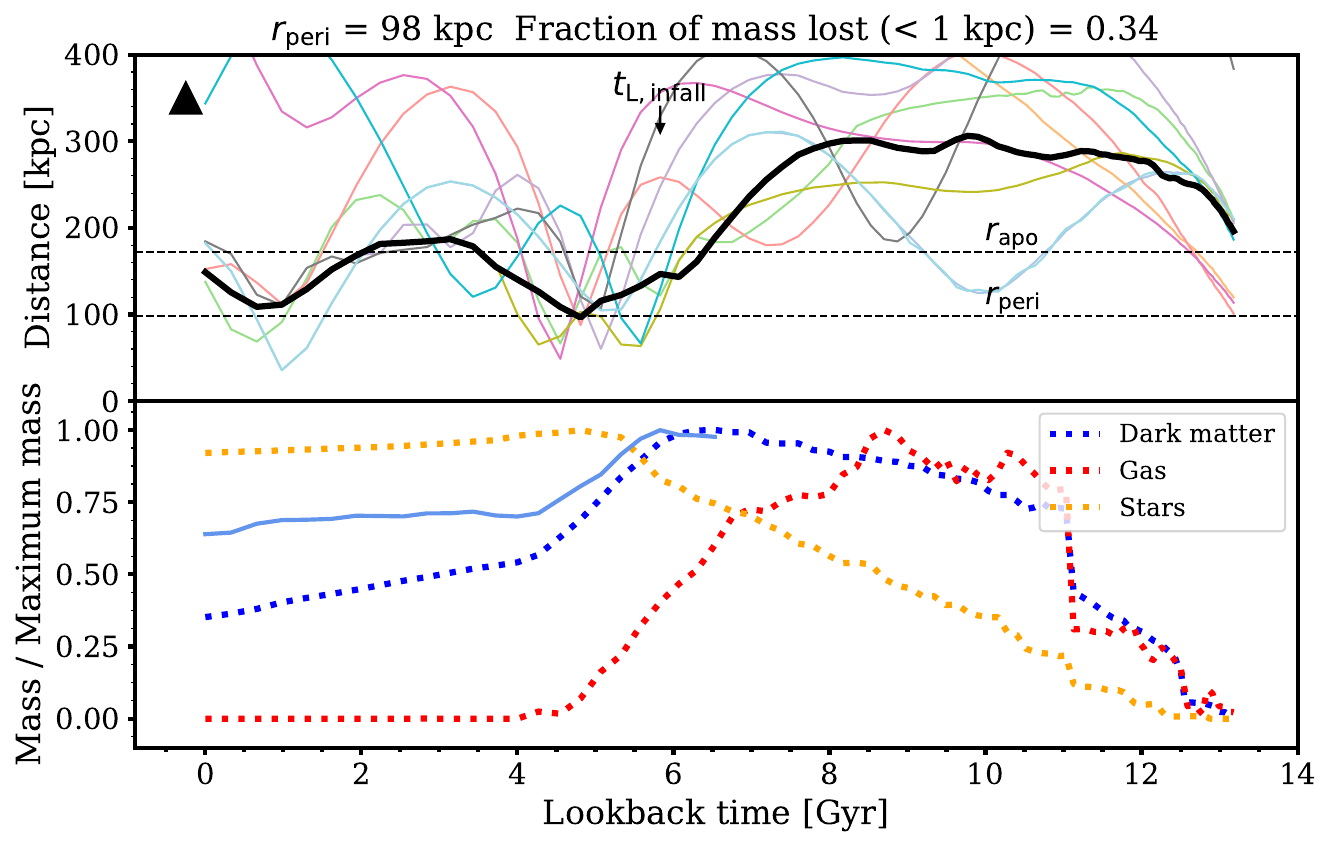}\par
  
  \includegraphics[width=\columnwidth]{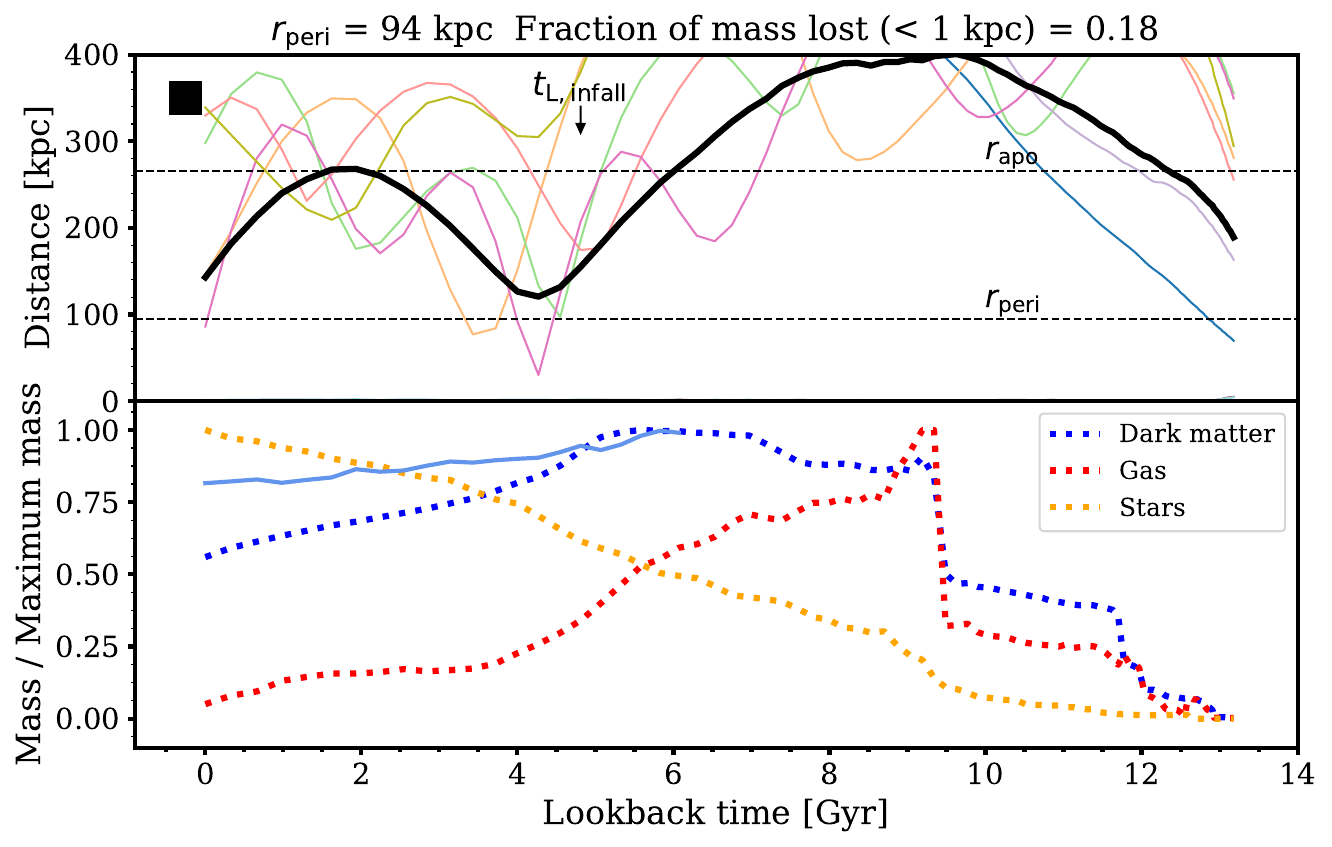}\par
  
    \end{multicols} 
    \caption{Individual formation and evolutionary histories of the 9
      Fornax analogues. Each analogue is marked by the corresponding
      symbol that we use throughout this work. The top subplots show
      physical distance to the host galaxy (thick solid line), as a
      function of lookback time. The inferred peri- and apocentres are
      shown as thin dashed lines. Lines of different colours show
      distances to other dwarf galaxies which, at any one point, were
      the nearest more massive galaxy than the Fornax analogue. The
      black arrow marks the infall time of each dwarf. The legend
      shows the inferred value of the orbital pericentre and the
      fraction of mass lost within 1~kpc. The bottom subplots show the
      bound dark matter (blue), gas (red) and stellar (yellow) masses,
      normalized by their historical maximum. The solid light blue line shows the evolution of dark matter mass within 1~kpc from the time the halo had its maximum mass.}
		\label{fig5}
	\end{figure*}

\begin{figure}

         \includegraphics[width=\columnwidth]{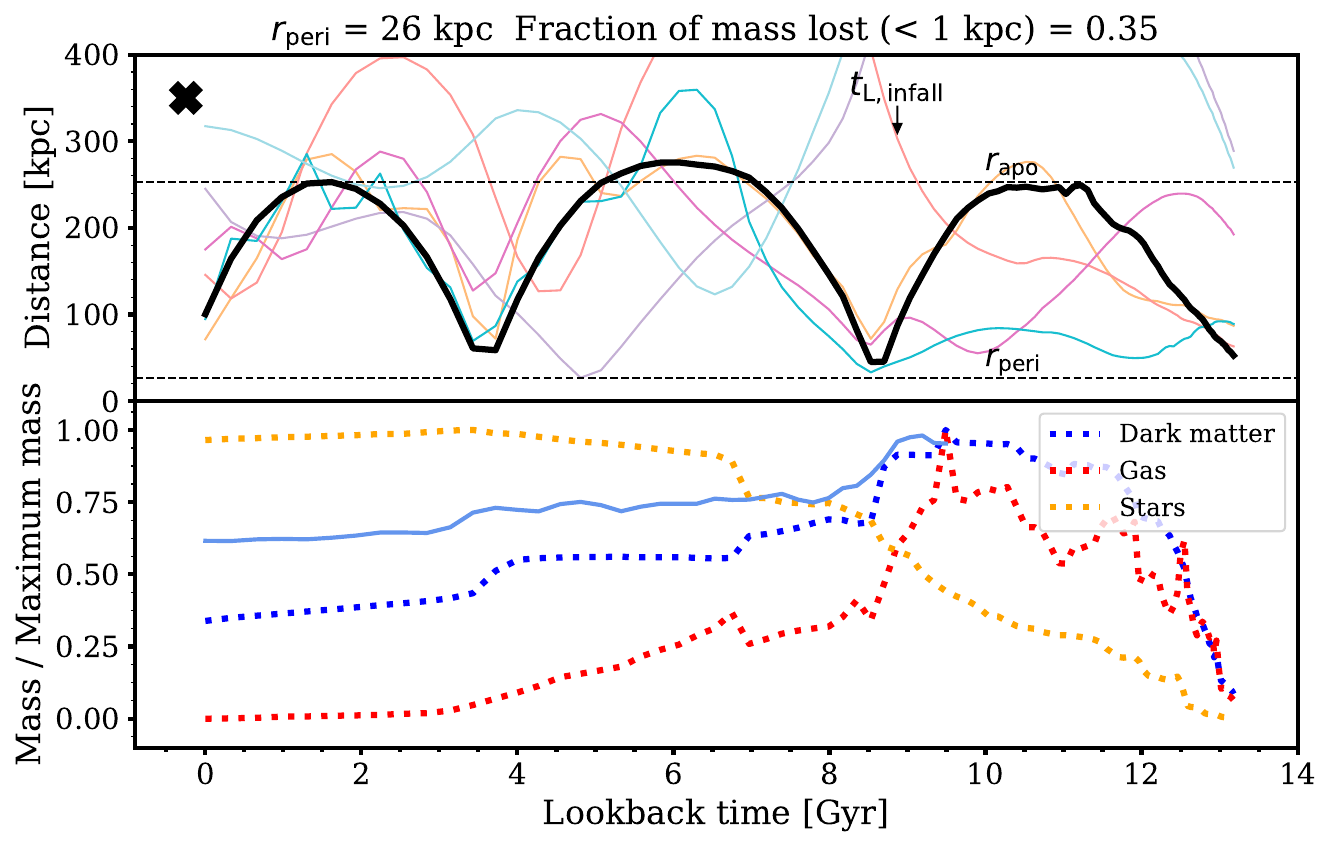}\par

         \includegraphics[width=1.\columnwidth]{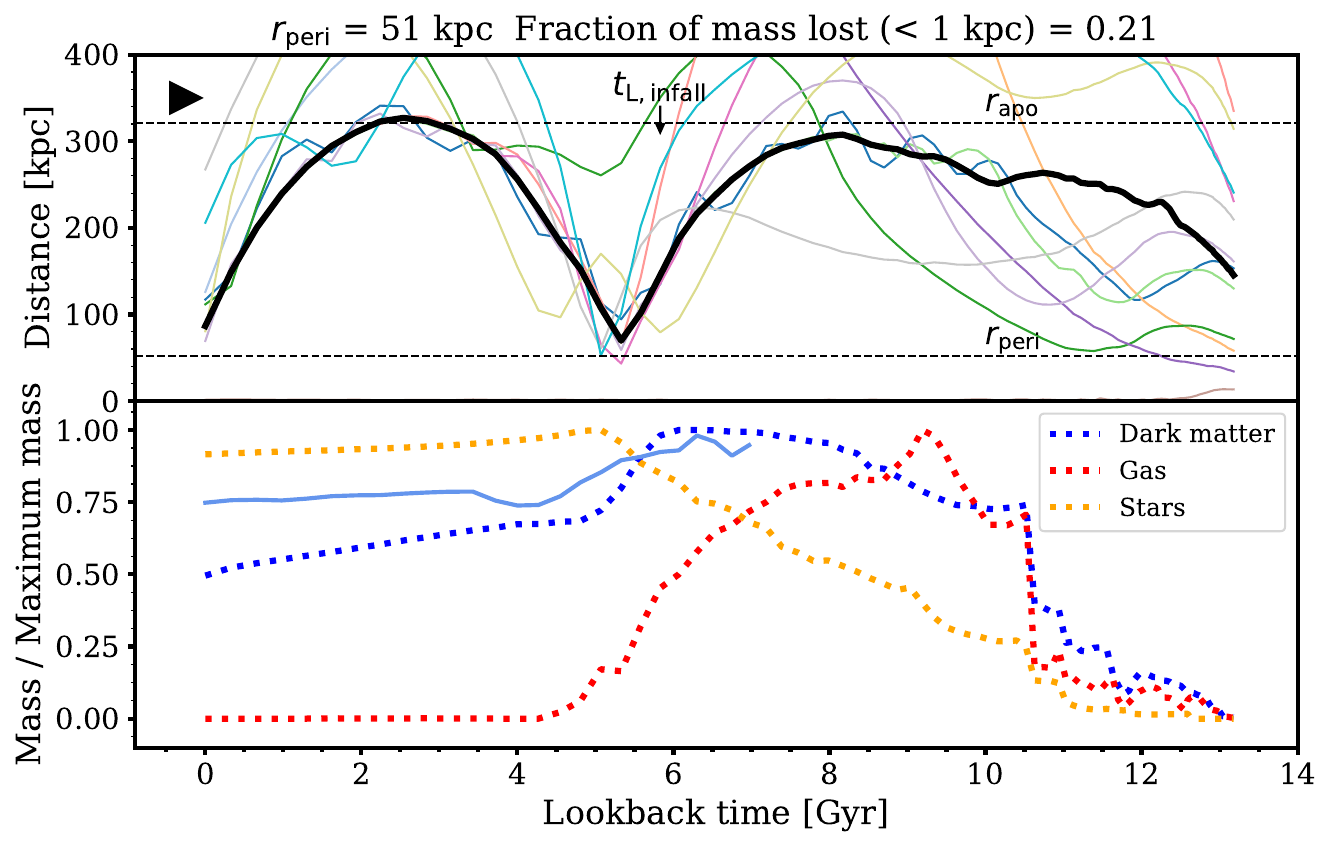} \par

		\includegraphics[width=1.\columnwidth]{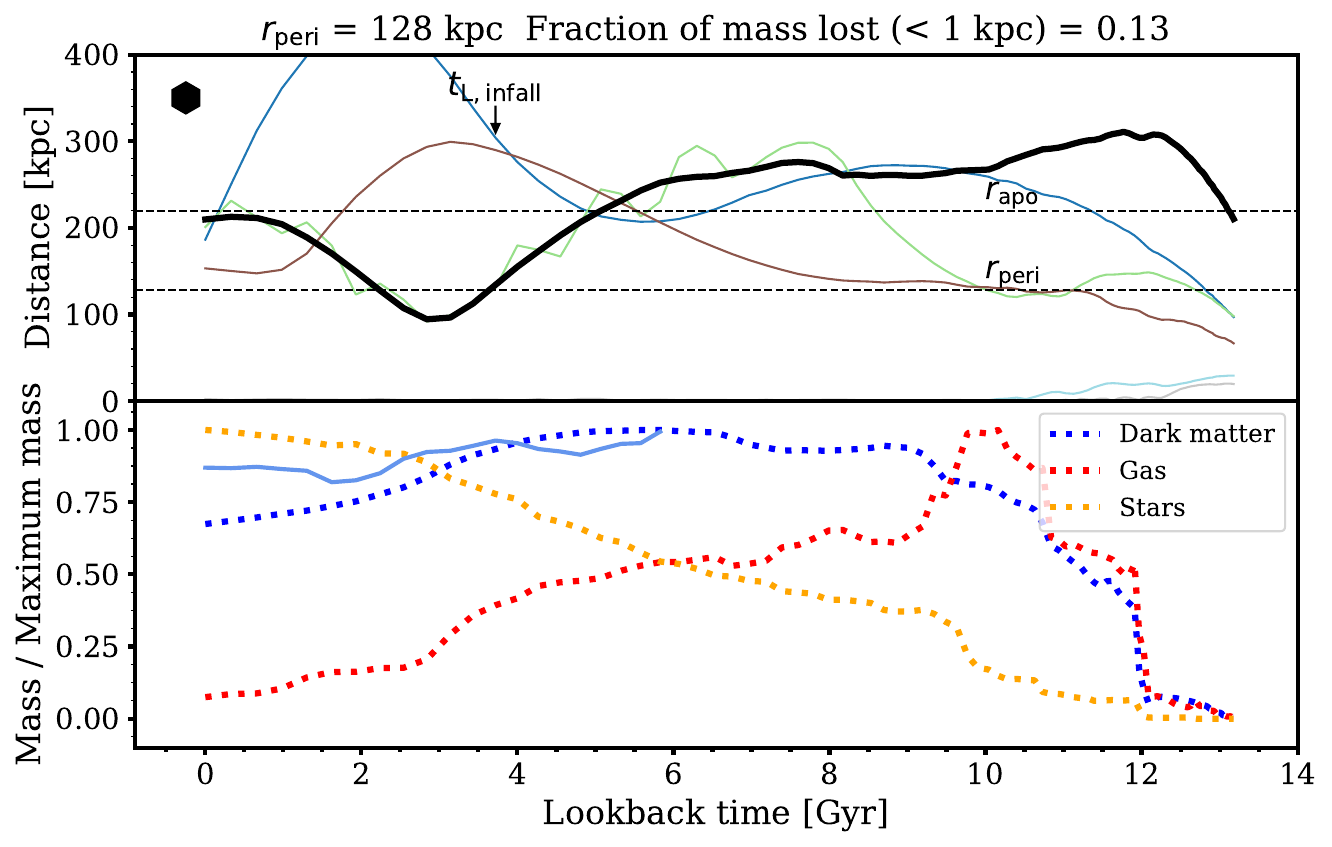}

		\contcaption{}

		\label{fig5cont}
	\end{figure}

\begin{figure*}
    \centering
   
    \includegraphics[width=2\columnwidth]{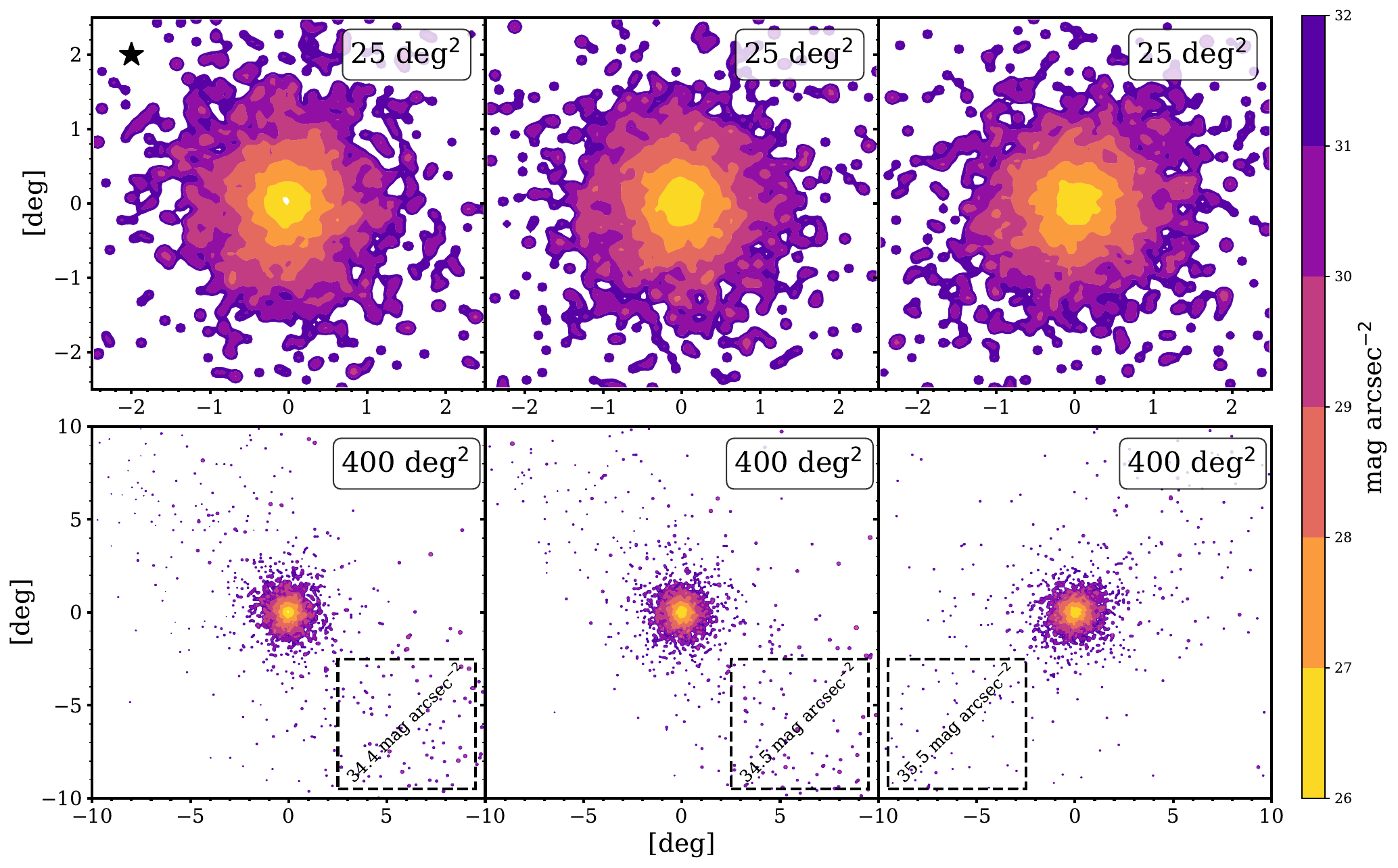}

    \caption{Surface brightness contour maps for our `star' Fornax, for survey areas of 25~deg$^2$ (as in \citealt{wangfornaxtidal}) and 400~deg$^2$, shown at the top and the bottom, respectively. Each pixel of 0.05~deg$^2$ has been smoothed using a Gaussian filter with a standard deviation of 0.05~deg. The dwarf is placed at a distance of 138~kpc and projected along the three axes of the simulation cube. The squares select an area of the stellar stream in which we compute a mean surface brightness of stream particles.}
		\label{fig6}
	\end{figure*}	

\begin{figure*}
    \centering

    \includegraphics[width=2\columnwidth]{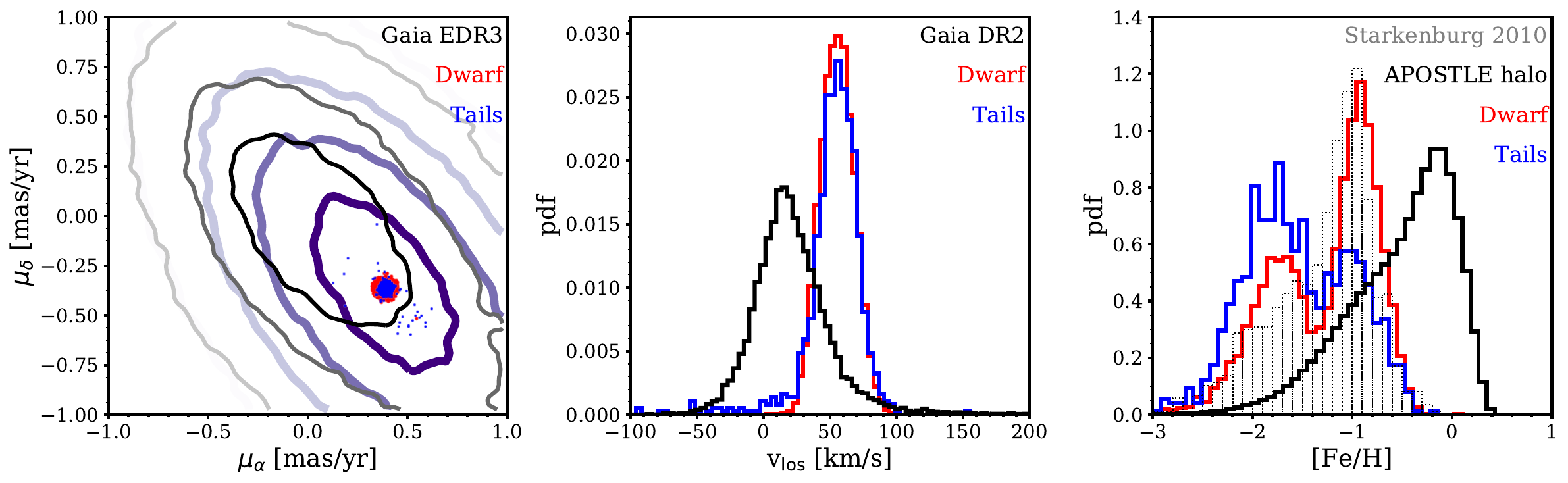}
    \caption{The chemo-kinematic properties of the `star' Fornax analogue. {\it Left:} The stellar proper motions from {\it Gaia EDR3} for stars within 1.5~deg of Fornax. The thick contours in shades of purple show regions of equal probability density. Our 'star' Fornax analogue is shown with red dots and its tidal tails are shown in blue. The proper motions of the analogue were centred on the Fornax values from \citet{gaia_edr3_pm}. The thin grey contours show regions of equal probability density for the stellar proper motions within the ``background" region $1.5<R~[\rm deg]<10$, centred on Fornax. The contours display the same probability density levels as for the $R<1.5~{\rm deg}$ region. The contours were smoothed with a Gaussian filter, with the same standard deviation in both cases, for clarity. In the computation of the contours, we include the errors on the proper motions and their correlations. {\it Middle:} the line-of-sight velocities of the `star' dwarf and of the tidal tails are shown in red and blue, respectively, and are centred on the mean line-of-sight velocity of Fornax \citep{gaia_edr3_pm}. The black histogram shows the distribution of line-of-sight velocities of "background" stars from {\it Gaia DR2} for the region $1.5<R~[\rm deg]<10$, centred on Fornax, with velocity errors included \citep{helmiGaia}. {\it Right:} the metallicity distributions of the `star' dwarf and the tidal tails, as inferred from our simulations, are shown with red and blue histograms, respectively. The dotted histogram shows the metallicity distribution of Fornax from \citet{starkenburg}. The black histogram shows the metallicity distribution of the stellar halo of the Milky Way analogue in APOSTLE, computed for the stellar particles located between 8~kpc away from the centre of the host galaxy and the virial radius, $R_{200}$.}
		\label{fig7}
	\end{figure*}	

We now explore the various mechanisms that lead to the formation of
our Fornax analogues. These are shown in Fig.~\ref{fig5}, where we have selected 9 analogues that are most compatible with Fornax in terms of $V_{\rm circ}(\rm{947~pc})$ and $M_V$. In the top subplots, the
thick black solid line shows the distance to the host galaxy. The
coloured lines show the distances from analogue Fornaxes to simulated dwarfs that at any given point in time were the nearest more massive galaxy. In
the bottom subplots, we show the evolution of the dark matter, gas and
stellar content. Note that for a number of dwarf galaxies there is a
steep rise in mass at early times ($\sim 11$~Gyr). This is likely
caused by {\sc hbt+} incorrectly identifying the main progenitor.

The majority of the dwarf galaxies have very similar formation
histories, in which their dark matter haloes are built up early on and
dark matter is subsequently lost after infall. Sudden drops in the
dark matter content can be seen when the dwarfs approach their
pericentres. The majority of our examples follow a very similar history of losing mass due to the effects of tidal stripping from the host galaxy following infall.

The `star' Fornax, our best Fornax match, is indeed a very peculiar example. Although this galaxy fell in only $\sim 4$~Gyr ago, it has been losing mass for nearly 9~Gyr. By redshift $z=0$, the galaxy has lost 80~per~cent of the maximum mass it once had and 47~per~cent of the mass within 1~kpc. This is because, on two separate occasions, the galaxy interacts with more massive dwarf galaxy fly-bys, each of which strip $\sim$25~per~cent of its initial
mass. While we would expect that the innermost dark matter particles would not be significantly affected in these interactions \citep{galacticdynamics}, some mass loss within 1~kpc is clearly visible. In this case, while the pericentre is indeed small (34 kpc), it is still not a good indicator of just how much mass the satellites has really lost.

The `thin diamond' Fornax is our second best Fornax analogue. This galaxy loses near 60~per~cent of its mass after two pericentres following its infall. The `thin diamond' is an excellent match to Fornax not only in terms of stellar mass and circular velocity but also in terms of quenching time ($\sim2$~Gyr ago) and its inferred infall time ($\sim8$~Gyr ago) derived via binding energy arguments \citep{rocha,wangfornaxinfall,fillingham_infall}.  This galaxy, together with the `cross' Fornax, are also clear examples of group infall (the orbits of companion galaxies are shown in blue). While group infall does not seem to play a role in the tidal stripping of these dwarfs, which can be seen to start occurring only after infall, we note it here as a plausible mechanism of `pre-processing' a satellite galaxy, prior to it crossing the virial radius of the host. \citet{santos_lmc} have shown that LMC-like objects are twice as likely in Local Group-like environments (14 out of 24 host galaxies in APOSTLE have an LMC analogue) than in isolated galaxies. Pre-processed satellites LMC-like objects can ultimately contribute to the dwarf population in the Milky Way.

The `thin diamond' Fornax has a pericentre of $\sim 60$~kpc, which helps explain the amount of tidal stripping occurring in this dwarf. The value of 60~kpc is well within the error bars for Fornax in a number of Galactic potentials (see Fig.~\ref{fig1}). Note, however, the mass loss in the `square` Fornax, which loses nearly 50~per~cent of its pre-infall mass and $\sim$25~per~cent of the mass within 1~kpc. This dwarf has a pericentre of $\sim$90~kpc, yet clearly it is able to lose a large fraction of its mass. Note also the `upwards triangle' analogue. The
pericentre here is 98~kpc; however, as the satellite falls in
$\sim6$~Gyr ago, we observe a drop in the dark matter mass by a factor
of 2. Around the same time, we can see that this galaxy is approached
by a series of more massive haloes that come as close as 50~kpc to
it. This accounts for the enhanced stripping in this dwarf. We see a similar, less severe, occurrence in the `right-point triangle' Fornax (Fig.~\ref{fig5}-{\it continued}), where the galaxy loses $\sim$40~per~cent of its initial mass. Also note that at the first pericentre, $\sim5$~Gyr ago in these galaxies, there is a burst in star formation. This is an example of one of the mechanisms for the formation of two distinct metallicity populations discussed by \citet{thedistinct}.

A complete outlier in terms of formation history is the `thick diamond' Fornax. This galaxy falls in only
$\sim 3$~Gyr ago and has lost less than 10~per~cent of its mass. This galaxy had undergone two minor mergers in the last 8~Gyr. This example is
slightly fainter than Fornax but has the same value of
V$_{\rm circ}(\rm{947~pc})$. Evidently, it started off as a very low mass halo that had built up to have a similar $V_{\rm circ}(\rm{947~pc})$ to Fornax. Tidal effects are clearly not important for this galaxy.

So far we have seen that, while the processes that lead to the
formation of Fornax analogues are quite similar (mass loss at
pericentre), $\frac{1}{9}$ of our best Fornax examples have lost some
dark matter mass prior to infall, $\frac{2}{9}$ have enhanced mass loss near pericentre by interactions with other dwarfs and $\frac{1}{9}$ came from a low mass halo that was assembled through late minor mergers and never
experienced significant mass loss.

\subsection{Has the tidal stripping of Fornax been observed?}	

From our analysis so far, it is clear that in order to match the present-day $V_{\rm circ}(\rm{947~pc})$ and $M_V$ of Fornax some tidal stripping is typically required (but not in all cases, as in the `thick diamond' example). Indeed, our best match to Fornax (the `star') has been severely tidally stripped, depleting this galaxy of both the dark matter and the stellar component. The eventful tidal history experienced by this dwarf has left traces in the form of tidal tails, potentially detectable by wide-field photometric surveys. The work of \citet{wangfornaxtidal}, using the 3-year data from the Dark Energy Survey (DES), have covered the 25~deg$^2$ area centred on the Fornax dwarf galaxy. Within this area, down to 32 mag~arcsec$^{-2}$, no tidal features were identified.

We shall now focus on the `star' Fornax, as this galaxy has the most mass in tidal tails of all of our examples. We collect all the stellar particles that have ever been bound to this dwarf and carry out a mock photometric survey. In Fig.~\ref{fig6}, we show the surface brightness maps for our `star' Fornax in 25~deg$^2$ and 400~deg$^2$ regions. As in \citet{wangfornaxtidal}, we use the pixel size of 3~arcminutes. The analogue Fornax is placed at a distance of 138~kpc. We use the small-angle approximation throughout. In order to compensate for the sampling noise in our simulations, we smooth the luminosity in each pixel using a Gaussian filter with a standard deviation of 3~arcminutes. For stellar particles within the tidal tails, however, this is not sufficient to compensate for inadequate stellar sampling. We therefore instead compute the mean surface brightness inside a 7~deg$^2$ area in the tails (as shown with dashed squares in Fig.~\ref{fig6}).

It can be seen that the detection of stars in the tidal streams requires a surface brightness limit of $\sim35-36$~mag~arcsec$^{-2}$ (note that in the leftmost two panels of Fig.~\ref{fig6} the stream is denser as it follows a path close to the line-of-sight), which has not been reached with DES \citep{wangfornaxtidal}. Moreover, the area of 25~deg$^2$ does not seem to provide sufficient coverage to observe the presence of tidal tails, which are clearly visible in the survey area of 400~deg$^2$ (bottom of Fig.~\ref{fig6}). We conclude that a larger spatial coverage and deeper surveys are needed for the detection of tidal tails in Fornax (provided it is stripped as severely as our `star' analogue) and their existence is not presently ruled out. 

It should be noted that in our analysis so far we have not considered foreground and background stars, which would significantly complicate the identification of stellar streams. This could in principle be resolved in the future with kinematic and proper motion studies. We explore these prospects further in Fig.~\ref{fig7}, where we show the proper motions, line-of-sight velocities and [Fe/H] values\footnote{The metallicities have been computed as in \citet{thedistinct}, assuming the solar mass fractions of 0.0014M$_{\odot}$ for iron and 0.7381M$_{\odot}$ for hydrogen.} for the stellar particles bound to our simulated dwarf (red), stellar particles in the tidal tails (blue) and the line-of-sight contaminants (black). 

In the left panel of Fig.~\ref{fig7} we display the proper motions from {\it Gaia EDR3} \citep{gaia_edr3} for two regions centred on Fornax: the region within 1.5~degrees of Fornax (purple contours) and the region between 1.5 and 10 degrees from the centre of Fornax (grey contours), corresponding to "background" stars. Note that the latter may indeed include the tidal tails of Fornax. The density contours are computed by sampling a 1000 times, for each {\it Gaia} source, a multivariate normal distribution defined by the errors on the measurements of the right ascension, $\mu_{\alpha}$, and declination, $\mu_{\delta}$, proper motions and their correlations. The contours show regions of equal probability density in both samples. The stellar particles belonging to our analogue Fornax are shown in red and its tails are shown in blue. Note that the purple contours are centred on the proper motion measurement of Fornax, where we have artificially placed our 'star' analogue.

It is clear that the proper motions of the tails cluster in the same region as the dwarf itself, potentially easing their detection in the future. However, as is clear from the purple probability density ellipses in Fig.~\ref{fig7} (which correspond to stars that are likely to be predominantly Fornax members), the errors on the proper motions are at present large (owing primarily to the relatively large heliocentric distance of Fornax) and it would be difficult to exploit the proper motion measurements to identify the stars belonging to the tidal tails in the near future.

The middle panel of Fig.~\ref{fig7} displays the line-of-sight velocities of the `star' Fornax (red) and its tails (blue), centred on the mean value for Fornax \citep{gaia_edr3_pm}. The black histogram shows the available line-of-sight velocities from {\it Gaia DR2} \citep{helmiGaia} for the region between 1.5 and 10 degrees away from the Fornax centre, where we have sampled the errors on each velocity measurement. Again, it is evident that the line-of-sight velocities of the stars in the tails peak at the same value as Fornax itself, potentially easing their identification.

On the right panel of Fig.~\ref{fig7}, we display the metallicity distribution of our `star' analogue Fornax in red and its tails in blue. Evidently, there exists a metallicity gradient in this galaxy, resulting in more metal-poor tidal tails. The metallicity distribution of the halo stars in the Milky Way analogue are shown with the black histogram. Here we have included stellar particles between 8~kpc from the centre of the main galaxy and out to the virial radius, excluding the ``disk" stars (all stellar particles born bound to the main galaxy) and stellar particles that belong to other satellite dwarfs at $z=0$. It can be seen that Fornax stellar particles, and the particles in the tails, in particular, peak at much lower metallicities than the halo of the simulated Milky Way analogue. This may, in the future, ease their detection.

It should also be noted that the metallicity distribution of our `star' Fornax analogue has a double-peaked shape (red histogram on the right of Fig.~\ref{fig7}), suggesting the presence of two distinct metallicity subpopulations. At least two metallicity subpopulations have also been previously identified in the Fornax dwarf spheroidal \citep{battagliafornax,letartefornax,amoriscoevans}. The possible origin of these populations has been discussed in detail in \citet{thedistinct}. For comparison, with the dotted grey histogram we show the metallicity distribution for Fornax inferred from spectroscopy \citep{battagliafornax,starkenburg,lemasle}. The two distributions show good agreement; however, we note that the stars outside the half-light radius of Fornax, which would be typically more metal-poor, are likely under-sampled. The shape of this metallicity distribution, with a dominant metal-rich peak, is reminiscent of simulated dwarfs which have formed their metal-rich population via a merger. The `star' Fornax has clearly undergone a merger $\sim11$~Gyr ago, triggering growth in stellar mass, which is subsequently also enhanced at the time of interactions with other dwarfs (see the yellow dotted line in Fig.~\ref{fig5}). The distinct metallicity populations in this Fornax analogue are also spatially segregated, resulting in tidal tails of low metallicity. 

\subsection{Dark matter density profiles of tidally stripped Fornax analogues}
\label{azicorrection}

\begin{figure*}
\centering
	\includegraphics[width=2\columnwidth]{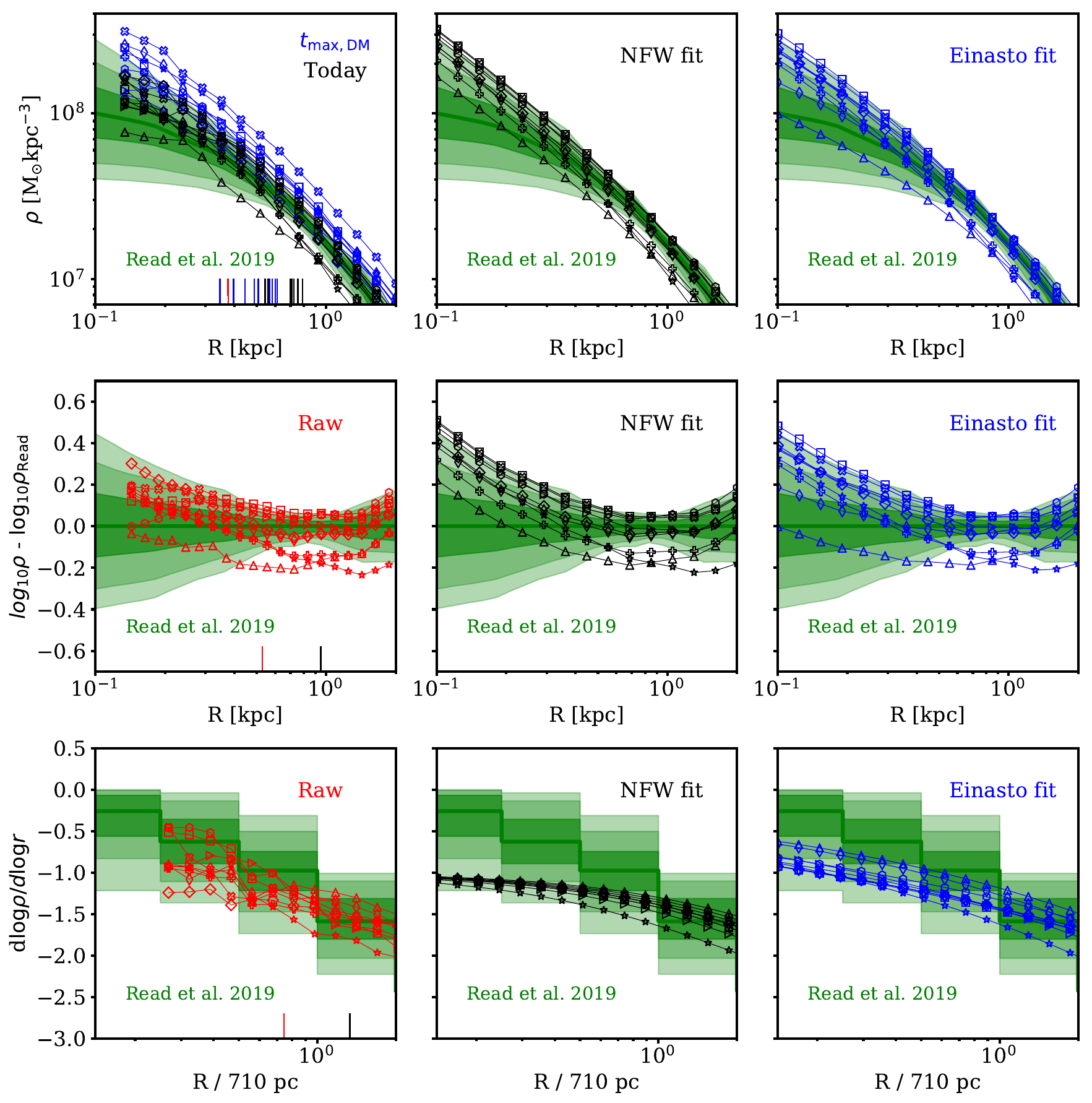}
	
    \caption{{\it Top left:} dark matter density profiles of
      our Fornax analogues, marked with their respective symbols, at
      the time when they had their maximum dark matter mass (blue) and
      at the present day (black). The green shaded bands are the 68, 95 and 97.5 confidence intervals for Fornax from \citet{heating}. The red
      vertical line marks 2.8$\epsilon = 380$~pc. The black and blue
      vertical lines mark the locations of the \citet{power} radius
      for each dwarf today and at the time of maximum halo mass,
      respectively. The left limit on the profiles is set at the gravitational softening $\epsilon =134$~pc. {\it Centre left:} logarithmic difference between the $z=0$ density profiles of our simulated subhaloes and the constraints from \citet{heating} (red). The red vertical line
      marks $2.8\epsilon$ and the black vertical
      line the mean location of the \citet{power} radius. {\it Bottom left:} the log-slope of the density for our Fornax analogues, compared to the log-slope constraints for Fornax from \citet{heating}. The values are shown at fixed fractions of the half-light radius of Fornax (710~pc). The red and black vertical lines mark $2.8\epsilon$ and the mean \citet{power} radius, respectively. {\it Middle column:} the best NFW profile fits to our Fornax analogues at $z=0$ (top);  the logarithmic difference from the results of \citealt{heating} (middle) and the log-slope of the profile (bottom). {\it Right column:} the same as the middle column, but for the best Einasto profile fits. Einasto shape parameters are in the range 2.96 - 4.54.}

\label{fig8}
\end{figure*}	

Now that we have established that the majority of our Fornax analogues
are able to lose dark matter from the inner 1~kpc due to tidal
effects, we wish to compare their dark matter density profiles to the profile
inferred for Fornax from its stellar kinematics by \citet{heating}. Their work used a high-order Jeans analysis method, {\sc GravSphere}, to infer the dark matter density distribution in Fornax. The validity of the method has been extensively tested in \citet{gravsphere}, \citet{readdraco} and \citet{tobeta}.

In the top left panel of Fig.~\ref{fig8}, we show the density
profiles of our Fornax analogues at the time they had their maximum
dark matter mass (blue) and at the present day (black). It is clear
that the inner density has diminished in these galaxies after
infall. A number of present-day density profiles appear
compatible within the 68-95$^{\rm th}$~per~cent confidence levels with
the result of \citealt{heating} (green shaded bands). However, there
are two important caveats. Approximately on the scale of
2.8$\epsilon\approx380$~pc (where $\epsilon$ is the gravitational
softening parameter), the dark matter profiles flatten due to resolution effects \citep{ludlowa}. Another important effect is the flattening of the density profiles due to collisional relaxation. A robust prediction for the convergence radius of density profiles
comes from \citet{power}, who show that NFW profiles are converged at
the radius where the collisional relaxation timescale,
$t_{\rm relax}\approx 0.6t_0$, where $t_0$ is the age of the
Universe. This corresponds to approximately 0.6~kpc for our dwarfs.

It is well established that haloes maintain their cuspy
shapes in higher-resolution dark-matter-only $N$-body simulations,
although with a larger concentration parameter than their field
counterparts \citep{aquarius, navarroAquarius,errani_reconstruct}. As
we are particularly interested in the density in the inner regions, we fit NFW profiles directly to the $z=0$ density profiles\footnote{We use the \textit{enclosed} densities for the fitting as they are significantly less noisy in simulations than their differential counterparts.} of our analogues between the \citet{power} convergence radius and the tidal radius\footnote{We define the tidal radius as the radius where the enclosed density of the satellite is equal to the enclosed density of the host up to that radius.}. The fitting is carried out using the \textsc{lmfit} algorithm \citep{lmfit}, minimizing the $\chi^2$ function in 21 log-density bins, as defined in \citet{neto}. The NFW profile is defined as 

\begin{equation}
 \rho(r) = \frac{\rho_s} {\frac{r}{r_s} \bigg(1 + \frac{r}{r_s}\bigg)^2  }.
 \label{nfw_equation}
\end{equation}

We find the scale density, $\rho_s$, and the scale radius, $r_s$, that minimize  $\chi^2$. The resulting profiles are shown in the top-centre panel of Fig.~\ref{fig8}, while the middle-centre panel shows the logarithmic difference of the fitted NFW profiles from the profiles inferred by \citet{heating}. It is clear that our Fornax analogues are typically denser in the centre than \citet{heating} predict. Only three analogues are consistent with \citet{heating} within 99.7~per~cent confidence in the region 0.1-2~kpc. At 150~pc, 6 of the analogues have densities consistent with Fornax within the 99.7~per~cent confidence interval and 3 within the 95~per~cent confidence interval. 

So far we have only compared the absolute magnitudes of the density at each radius. In order to compare the
{\it shapes} of the profiles more effectively, we display the log-slope of the density in the bottom row of Fig.~\ref{fig8}. The red lines show the
slopes obtained directly from the simulations and the black lines
the slopes of the fitted NFW profiles at each radius. The green bands show the results of the kinematic analysis of \citet{heating}. Note that \citet{heating} use a broken power-law fit to the density profiles, where the breaks occur at fixed intervals of the half-light radius $[0,0.25,0.5,1,2,4]R_e$. It can be seen that raw simulation profiles with shallow cusps on the scale of the gravitational softening (red vertical line in Fig.~\ref{fig9}), are typically consistent with the result of
\citet{heating}\footnote{We note that the broken-power-law model for the density distribution in {\sc GravSphere}, together with the requirement that the density should decrease outwards, generally disfavours density distributions with cores $>0.25 R_e$ \citep{heating}. We have checked, however, that the {\sc coreNFWtides} model, where such cores are allowed, still permits NFW-like cusps in the central regions of Fornax within the 95~per~cent confidence levels.} within the 95~per~cent confidence level at all radii, while the NFW profiles are consistent within the 97.5~per~cent confidence level in the central regions and within 95~per~cent near the half-light radius.

We additionally consider the Einasto functional form \citep{navarroAquarius} for the simulated Fornax analogues,
\begin{equation}
 \rho(r) = \rho_{-2} \exp {\left[ -2\alpha \left( \left(\frac{r}{r_{-2}}\right)^{\frac{1}{\alpha}} -1 \right)\right]  },
 \label{einasto_equation}
\end{equation}
with $\rho_{-2}$ being the density at the radius $r_{-2}$, where the log-slope of the density profile is -2. The parameter, $\alpha$, is the Einasto shape parameter. Again, we find a set of parameters $(\rho_{-2}, r_{-2}, \alpha)$ that minimize $\chi^2$. We find Einasto shape parameters in the range 2.96~-~4.54, consistent with the range found in \citet{dicintio} for cuspy subhaloes in the DMO and SPH versions of the CLUES simulations. 

The results of the fit are shown in the right column of Fig.~\ref{fig8}. Einasto profiles allow for inner density cusps that are shallower than NFW. With this parameterization, we find that 5 of the Fornax analogues are consistent with the limits of \citet{heating} within the 97.5~per~cent confidence interval in the radial range 0.1-2~kpc. At 150~pc, 8 of the analogues have densities consistent with Fornax within the 97.5~per~cent confidence level and 4 within the 95~per~cent confidence level. In terms of the log-slope of the density (bottom right panel of Fig.~\ref{fig8}), 2 of the analogues are consistent with Fornax across the radial range 0.1-2~kpc and, as in the NFW case, all profile shapes are consistent with Fornax within the 97.5~per~cent confidence levels. 

It is clear from Fig.~\ref{fig8} that central dark matter densities
compatible with Fornax can be obtained through the effects of tidal
stripping; however, based on our extrapolation, the shapes of the
corrected density profiles of stripped NFW haloes appear to be less
similar to that of Fornax than those of haloes with a shallow cusp or
a small core in the central regions. However, we also note that the relatively small mass of stellar disks in APOSTLE host galaxies suggests that the strength of tidal effects is underestimated in these simulations, particularly for small-pericentre orbits. In the left panel of Fig.~2, the enhanced stripping would result in lower values of $V_{\rm circ}$, accompanied by little change in $M_V$. This could place the central densities of simulated analogues in better agreement with \citet{heating}.

\subsection{The tidally induced reduction in the central densities of Fornax analogues}
\label{heating_comp}

\begin{figure*}
 \begin{multicols}{2}
\includegraphics[width=0.88\columnwidth]{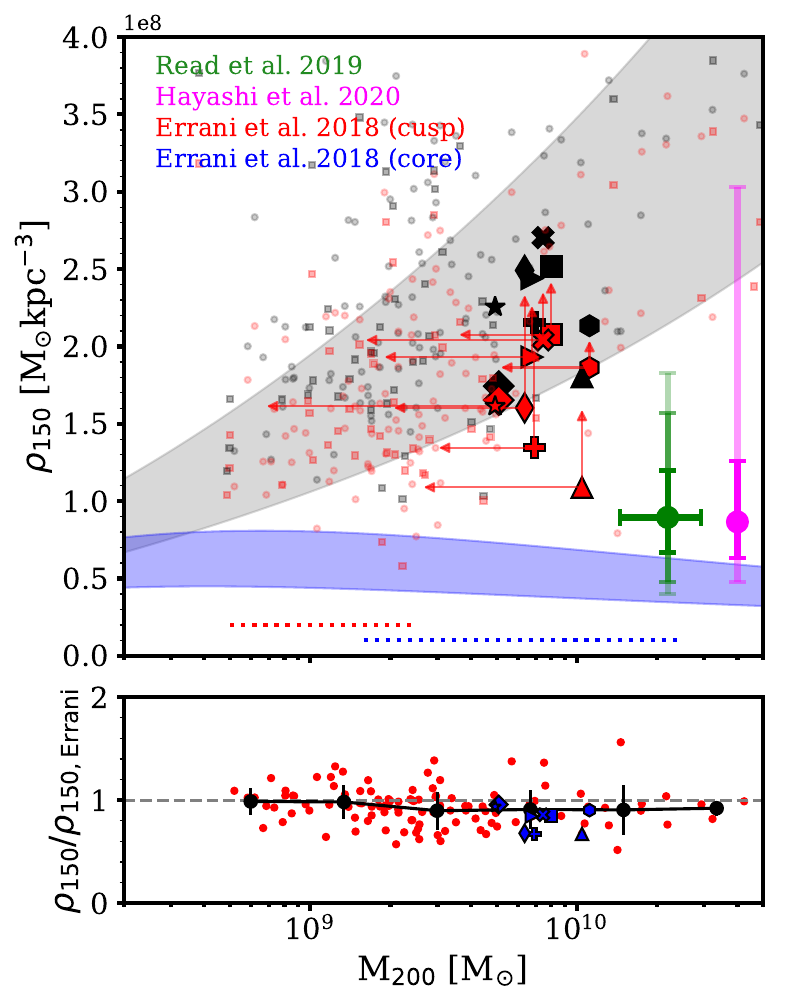}

\includegraphics[width=\columnwidth]{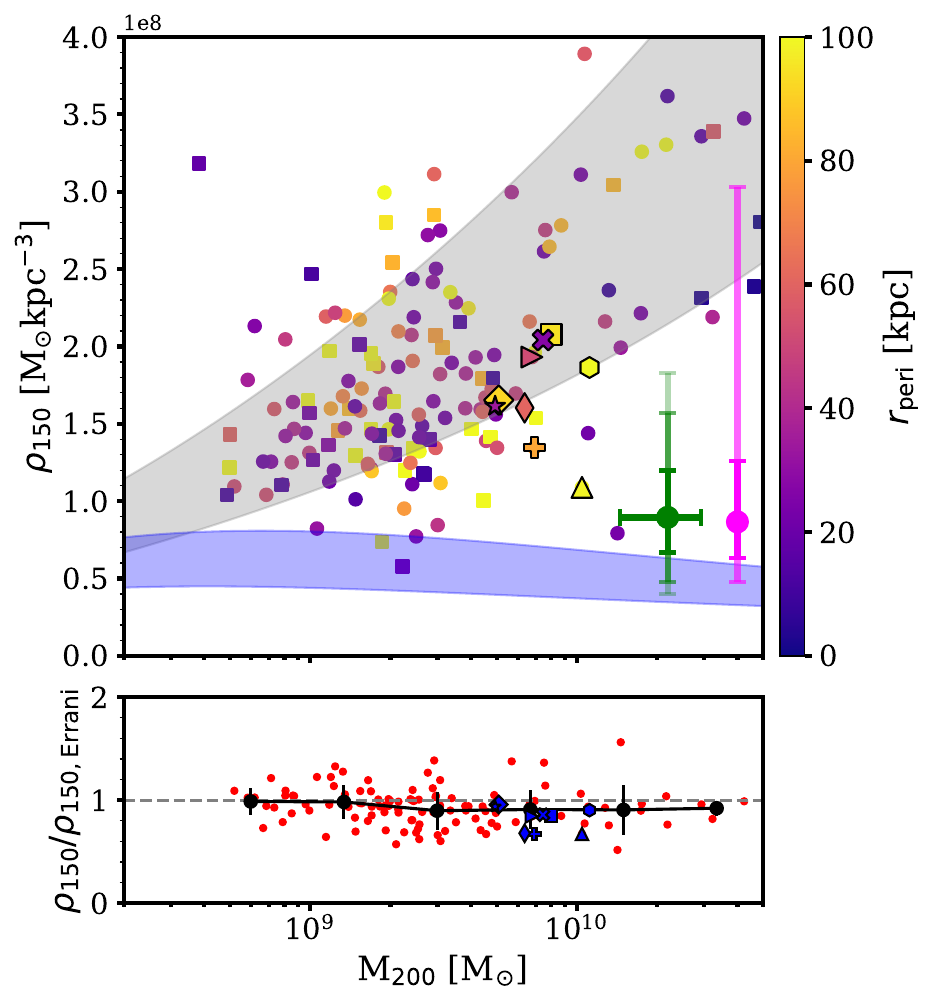}
   \end{multicols}

   \caption{{\it Left:} the density at 150~pc,
     $\rho_{150}$, as a function of all-time maximum virial mass,
     $M_{200}$, of each dwarf satellite. The black symbols show the
     densities at the time of maximum mass within $R_{\rm max}$ and the red symbols the
     densities at the present day. Circles show the densities inferred from a direct NFW profile fit, while squares show densities inferred from application of the \citet{errani_navarro} tidal tracks to the NFW profile fits at the time of maximum mass within R$_{\rm max}$. The green point and error bars
     display the median and 68-95-97.5 confidence intervals of $\rho_{150}$
     for Fornax from \citet{heating}; the point is positioned at the pre-infall halo mass value derived by \citet{readerkal}. The pink errorbar shows the 68 and 95~per~cent confidence intervals of $\rho_{150}$ for Fornax, derived by \citet{hayashi}. The pink point is offset to the right of the green one for clarity. The dotted red and blue horizontal lines show the 1$\sigma$ span of the total present-day mass of Fornax for a cuspy (red) and cored (blue) halo, derived by \citet{errani}. The grey shaded region is derived from the mass-concentration relation of \citet{mass-conc}
     and the blue shaded band is the equivalent relation for CDM
     haloes that form a core below the half-light radius.  The various
     symbols mark our 9 Fornax analogues. Each of these have $\rho_{150}$ values derived by fitting an NFW profile. The red upwards-pointing arrows show their $\rho_{150}$ values derived using the \citet{errani_navarro} tracks. The left-pointing arrows show the total bound mass at $z=0$. The bottom panel shows the comparison between the $z=0$ values of $\rho_{150}$ derived from an NFW profile fit (where the fit was possible) and the densities derived from the method of \citealt{errani_navarro} (red dots). Blue symbols are our Fornax analogues. The two estimates are consistent with each other for $M_{\rm 200}<3\times10^{10}M_{\rm \odot}$ within 1$\sigma$, though there is a systematic bias for $M_{\rm 200} > 10^9 M_{\rm \odot}$ of $\sim10$~per~cent. {\it Right:} the values of
     the present-day $\rho_{150}$ and all-time maximum, $M_{200}$,
     coloured by the inferred orbital pericentres of each
     dwarf satellite. Different symbols highlight our Fornax analogues. }
\label{fig9}
\end{figure*}

We now explore how the dwarf galaxies in our sample compare to Fornax
in the context of the central density - virial mass ($\rho_{150}$ - $M_{200}$) relation (see the middle panel of fig.~5 in \citealt{heating}). To obtain the $\rho_{150}$ values, we fit NFW profiles to the subhaloes in our simulations at $z=0$, as described in Section~\ref{azicorrection}. The estimated densities at 150~pc are shown as red circles in the left panel of Fig.~\ref{fig9}. We also show the densities of these subhaloes at the time of maximum $R_{\rm max}$, where $R_{\rm max}$ is the radius where the circular velocity profile peaks\footnote{For pre-infall haloes, we fit the NFW profile in the radial range between the \citet{power} convergence radius and $R_{200}$ in 31 logarithmic bins.}. The time of maximum $R_{\rm max}$ corresponds approximately to the infall time of a subhalo. In certain cases, an NFW profile provides a poor fit\footnote{Here we define the goodness of fit as the fitted profile being within 15~per~cent of the raw profile across the entire fitted radial range.} to the subhaloes at $z=0$. For these subhaloes, we apply the tidal tracks of \citet{errani_navarro} (using their equations 7 and 9) to the fitted profile at the time of maximum $R_{\rm max}$ to deduce the value of $\rho_{150}$ at $z=0$. These values are shown with red squares, while the black squares are the $\rho_{150}$ values at the time of maximum $R_{\rm max}$, obtained through fitting NFW profiles. In the bottom subplot of Fig.~\ref{fig9} we compare these estimated values to the ones obtained from NFW profile fits for cases where such a fit is possible. It can be seen that these estimates are effectively unbiased at all halo masses. We further discuss the \citet{errani_navarro} tracks and their applicability to our simulations in Appendix~B.

The black and red symbols of different shapes represent our Fornax analogues. The grey shaded band is taken from the mass-concentration relation of \citet{mass-conc} and the blue band is the densities expected for dwarfs with the same masses and concentrations that have undergone complete core formation below the half-light radius (the corresponding density profile is
described in \citealt{alltheway}). The green point with error bars is the inferred density of
Fornax at 150~pc from \citet{heating}, while the pink point shows the value inferred by \citet{hayashi} using axisymmetric Jeans analysis. In this figure, we notice two interesting features: 

Firstly, the pre-infall $M_{200}$ for our
sample of Fornax analogues is systematically lower than that inferred
by \citet{readerkal}. We predict a pre-infall
$M_{200}\approx 4-9 \times 10^9 M_{\odot}$ for Fornax. The origin of this discrepancy is not straightforward and we will discuss this point further in Section.~\ref{sec:fornaxmass}. The dotted red and blue lines show the $1\sigma$ span of the present-day masses of Fornax inferred by \citet{errani} for cuspy and cored haloes, respectively. As the horizontal axis of Fig.~\ref{fig9} displays the pre-infall halo mass, we show the present-day bound mass for our Fornax analogues with left-pointing arrows. Five of the analogues are consistent with the value inferred by \citet{errani} within 1$\sigma$.

Secondly, there is a large number of dwarf satellites, including several of our
dwarf analogues, whose $\rho_{\rm 150}$ is consistent with that of Fornax \citep{heating}, purely due to tidal effects. It can be seen that tidal effects in our simulations are able to reduce the central density of satellite dwarfs by a factor of 2-3. We also note that all of the analogues are consistent with the $\rho_{150}$ value inferred by \citet{hayashi} within $2\sigma$. 

An important aspect of the argument presented by \citet{heating} is
that the classical dwarfs shown in their fig.~5, including Fornax, are
unlikely to have undergone significant tidal stripping and shocking
based on their orbits as inferred from {\it Gaia} data. Thus, the low
density in Fornax is interpreted as being due to core formation rather
than tides. In the right panel of Fig.~\ref{fig9}, we show the
present-day densities of dwarfs as a function of $M_{200}$, coloured
by their inferred pericentres. We emphasize that these pericentres are
derived using a directly measured spherically symmetric host halo
potential from the simulations. 

The majority of dwarfs with particularly low central densities do
indeed have low pericentres; yet there are a number of outliers. Most
notably, our `upwards triangle' Fornax, with a pericentre of 98~kpc, has a
density comparable to that of Fornax. We do, however, note that,
unlike Fornax, this galaxy was quenched $\sim 5$~Gyr ago. The `plus' Fornax analogue is another example with a pericentre of
79~kpc and a central density compatible with that of Fornax. This
galaxy was only quenched $\sim$2~Gyr ago. Its halo mass, however, places it
in a position, where it is fully compatible with ${\rm \Lambda CDM}$.  We note again that a number of Milky Way
potentials in the literature predict pericentres for Fornax as low as
20-30~kpc.

\section{How likely is the tidal origin of the low dark matter density in Fornax?}
\label{sec5}

We have so far shown that in $\Lambda{\rm CDM}$ simulations in which baryonic feedback {\it does not} cause core formation, `core-like' densities may be encountered in Fornax-like haloes with eventful tidal histories which are not always reflected in their inferred orbital pericentres. There are four main possible sources of tension between our results and previous works: {\it i)}~high central densities observed for dwarf galaxies with smaller pericentres than Fornax \citep{kaplinghat}; {\it ii)}~the value of the pre-infall mass, $M_{200}$, of Fornax; {\it iii)}~the assumed pericentre of Fornax; {\it
  iv)}~possible numerical effects. We discuss these below.
  
\subsection{Comparison with the density-pericentre anticorrelation}  
  
 \begin{figure*}
    \centering
		\includegraphics[width=2\columnwidth]{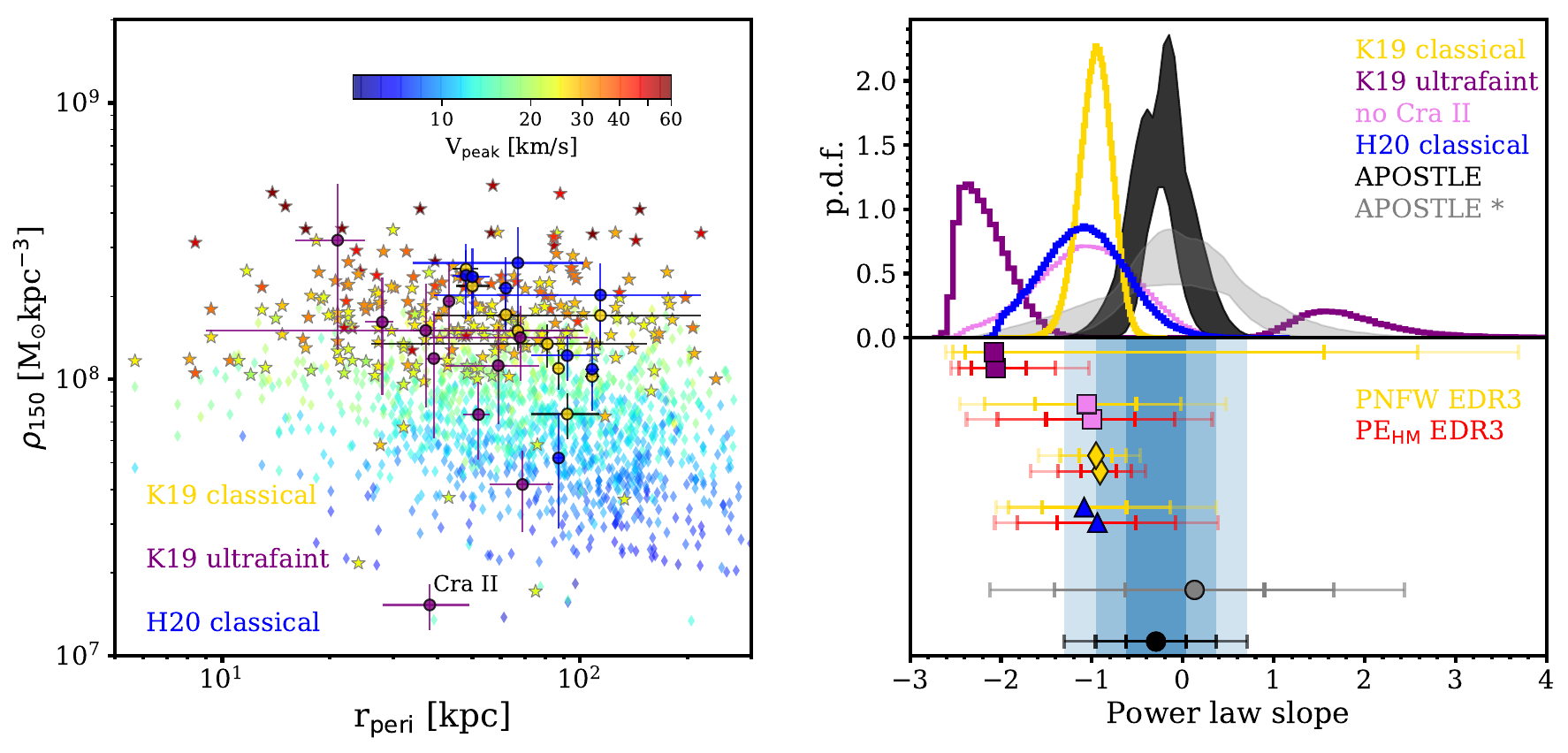}

		\caption{{\it Left:} the density at 150~pc as a function of pericentric distance, for the entire population of resolved satellites in APOSTLE. Satellites hosting at least one stellar particle are identified with `star' symbols and dark satellites with `diamonds'. The symbols are coloured by their historical maximum circular velocity, $V_{\rm peak}$. Haloes with $V_{\rm peak}\gtrsim19$~km/s form stars. Circles with error bars show the densities at 150~pc inferred by \citet{kaplinghat} for 9 of the Milky Way's classical dwarfs, with  pericentres taken for the {\it PNFW} Milky Way potential using {\it Gaia EDR3} \citep{gaia_edr3_pm}. Blue circles are density estimates for the same dwarfs (apart from Canes Venatici I) by \citet{hayashi}. The purple circles show the densities of the sample of ultra-faint dwarfs from \citet{kaplinghat}. {\it Top right:} the distribution of best-fit power-law slopes for the \citet{kaplinghat} sample of classical dwarfs (yellow), the \citet{kaplinghat} sample of ultra-faints (purple), ultrafaints excluding Crater~II (pink), and for the \citet{hayashi} sample of classical dwarfs (blue). The entire sample of satellites for each Milky Way analogue in APOSTLE is shown in black, where the bands bracket the distributions. The APOSTLE satellites of each analogue which host at least one stellar particle are shown in grey. {\it Bottom right:}  median (symbols) and 68, 95 and 97.5 confidence intervals (error bars) for the power slopes, computed for each sample of dwarfs in the {\it PNFW} (yellow) and the {\it $\rm PE_{\rm HM}$} (red) potentials \citep{gaia_edr3_pm}. For APOSTLE, we show the mean of the power-law slope distributions for the entire sample of satellites (black) and for the sample of luminous satellites (grey). The blue bands highlight the 68, 95 and 97.5 confidence intervals of the mean power-law slope for all satellites in different Milky Way analogue potentials. }

		\label{fig10}
	\end{figure*} 
  
  A possible discrepancy with our results is presented by \citet{kaplinghat} who have shown that the central densities of a sample of 9 bright Milky Way satellites appear to increase with decreasing orbital pericentre. It therefore seems paradoxical that Fornax, with its relatively large orbital pericentre, should have its density significantly reduced by tides, while other dwarfs, with smaller inferred pericentres, do not.
  
  Firstly, we emphasise that for our best Fornax analogue much of the stripping occurs {\em prior to infall} into the host galaxy. In another example (the galaxy indicated by an `upward triangle' in some of the figures), the pericentre is, in fact, $\sim100$~kpc, but stripping is enhanced by encounters with other massive objects. Secondly, as we have pointed out in Fig.~\ref{fig1}, the pericentre of Fornax is dependent on the assumed Milky Way potential. Note that this is not the case for many other classical dwarfs \citep{asya}, mainly affecting those at large Galactocentric distances (Leo~I, Leo~II, Canes~Venatici~I).   

We now aim to establish whether a density-pericentre anticorrelation exists in our simulations. In CDM, this may occur because of tidal effects, whereby subhaloes with small pericentres are disrupted and those with large pericentres survive. This `survivor bias' would then lead to an anticorrelation in the density-pericentre plane \citep{kaplinghat}. In the left panel of Fig.~\ref{fig10} we show the densities at 150~pc of all satellites in APOSTLE (those with and those without stars) as a function of pericentre. We derive the values of $\rho_{150}$ by directly fitting an NFW profile to the subhaloes at the time of their maximum $R_{\rm max}$ and applying the method of \citet{errani_navarro} to estimate $\rho_{150}$ at the present day. As we have shown in Fig.~\ref{fig9}, this method is an unbiased estimate of the densities we would have obtained by a direct NFW profile fit.  

We show the densities estimated by \citet{kaplinghat} for a sample of classical dwarfs with yellow circles on the left of Fig.~\ref{fig10} and for the ultra-faint dwarf sample in purple. We also show the densities estimated for the same sample of classical dwarfs (excluding Canes Venatici I) by \citet{hayashi} with blue circles. In all cases, we assume the $PNFW$ Milky Way potential with {\it Gaia EDR3} data \citep{gaia_edr3_pm}.

The satellites in Fig.~\ref{fig10} are coloured by their value of $V_{\rm peak}$. Firstly, we see that dwarfs that are denser at the present day have, unsurprisingly, larger values of $V_{\rm peak}$. Secondly, we observe a trend whereby starting from a pericentre of $\sim$~100~kpc and below, the number of surviving subhaloes decreases for smaller $V_{\rm peak}$. This is similar to the `survivor bias' also seen in the dark-matter-only simulations presented by \citet{hayashi}. Some dwarfs with smaller $V_{\rm peak}$ clearly still exist and have small pericentres. However, we have verified that many of these are recently accreted objects. Finally, at very small pericentres ($<20$~kpc), a cluster of points can be identified, where the densities seem to decrease with pericentre. This is indicative of dwarfs that are currently in the process of tidal disruption.

In order to quantify the density-pericentre relation in our simulations, we repeat the procedure of \citet{kaplinghat}, now fitting a power law to the full population of subhaloes of each Milky Way analogue. The distributions of best-fit slopes are shown with a black histogram in right-upper panel of Fig.~\ref{fig10}, where the bands bracket the results for all satellite populations. The distribution of best-fit slopes for the classical dwarfs using the density estimates of \citet{kaplinghat} is shown as a yellow histogram and the distribution obtained using the density estimates of \citet{hayashi} is shown as the blue histogram. From the right-bottom panel of Fig.~\ref{fig10} it can be seen that these two distributions are statistically consistent with APOSTLE, within 1$\sigma$ for the \citet{hayashi} sample and 2$\sigma$ for the \citet{kaplinghat} sample, although a shallower anti-correlation is preferred in our simulations. Applying the same fitting procedure to all satellites hosting at least one stellar particle, we obtain the grey histogram in the bottom panel of Fig.~\ref{fig10}. This, in general, suggests no correlation for the sample of luminous satellites, although the distribution appears somewhat double-peaked, where some analogues tend to a weak negative correlation and some to a weak positive correlation. We must bear in mind, however, that these objects are a biased subset of the broader distribution with a generally negative trend (which includes ultra-faint dwarfs whose stellar populations are unresolved).

The purple histogram in Fig.~\ref{fig10} shows the distribution of inferred power-law slopes for the sample of ultra-faints from \citet{kaplinghat}. This generally prefers much steeper slopes than the classical dwarfs. One can also see a second peak in this distribution, at a positive power-law slope of 1.5. We show that this is primarily due to the inclusion of Crater~II in this sample -- a clear outlier in terms of its density. It may seem surprising that this dwarf should affect the inferred power-law slope to such an extent; however, this dwarf has the smallest error bars on the density of the entire sample of ultra-faints, giving this data point greater weight. If this dwarf is removed from the sample, the inferred power-law slope for the ultra-faints becomes consistent within 1$\sigma$ with the classical dwarfs (pink histogram).  

Although the distributions of the best-fit power-law slopes in the observations and simulations are consistent within 1 or 2$\sigma$, it is nonetheless clear that the data (both for the ultra-faints and the classical dwarfs) prefer steeper power-law slopes and, for the \citet{kaplinghat} sample, the scatter is narrower than in our simulations. There are two points to consider here: {\it i)}~the strength of the Milky Way potential and {\it ii)}~the sample of observational data.

On the first point, we must note that the stellar disks of the host galaxies in the highest-resolution APOSTLE simulations have relatively low masses. The stellar mass is $\sim1-2\times10^{10}M_{\odot}$ \citep{fattahi}, compared to the estimated $5\times10^{10}M_{\odot}$ for the Milky Way \citep{milkywaymass}. Comparing APOSTLE with the AURIGA \citep{auriga} simulations ($M_*\sim 1-7\times10^{10}M_{\odot}$), \citet{richings_auriga} have shown that only $\sim$~30~per~cent of subhaloes in AURIGA with halo mass $M_{\rm DM} > 10^8 M_{\odot}$ survive at a radius of 10-20~kpc (compared to dark-matter-only simulations), while in APOSTLE $\sim$~50~per~cent of the substructures survive. This is due to the stellar masses of the central galaxies in AURIGA being twice as large as those in APOSTLE. The Milky Way analogues in APOSTLE typically have lower stellar masses than the estimates for our Galaxy \citep{fattahi}, while the stellar masses in AURIGA are typically higher. We thus expect that the gradient representing surviving subhaloes in the density-pericentre plane would steepen in simulations where the mass of the stellar disk is higher.

On the second point, our simulated samples clearly include objects that are currently in the process of tidal disruption, the dwarf irregulars and the recently accreted objects. The sample of \citet{kaplinghat}, on the other hand, excludes Sagittarius dSph (with a pericentre of 15~kpc and likely a lower dark matter density than expected from the anti-correlation relation at this pericentric distance), as well as the Large and Small Magellanic Clouds \citep{lokas_sag, vasiliev_sag}. The ultra-faint sample excludes dwarfs which likely fell in as part of a group \citep{patel}, dwarfs where the sample of kinematic tracers is too small and dwarfs for which the measured velocity dispersion is too low to be inconsistent with no velocity dispersion (for example, Triangulum~II and Segue~II \citep{triangulumii,segueii}, both of which are very underdense galaxies with small inferred pericentres \citealt{gaia_edr3_pm}). 

Moreover, if one considers instead the central densities inferred for the classical dwarfs by \citealt{hayashi} (shown with blue points in Fig.~\ref{fig10}), using axisymmetric Jeans modelling (we note again that many of the classical dwarfs are known to be aspherical), a density-pericentre trend is no longer obvious and the power-law fit to these data is not necessarily justified. As the models of \citet{hayashi} are more permissive of different tracer distributions in 3D, the density error bars increase and several of the classical dwarfs have densities offset from those of \citet{kaplinghat} by more than 1$\sigma$. This suggests that the error bars of \citet{kaplinghat} lack a systematic error contribution due to the asphericity of the stellar and halo components. Additionally, as we have shown in this work, the inferred pericentres are dependent on the chosen Milky Way potential, such that the error bars on the pericentre are likely underestimated, particularly for the more distant dwarfs. 

The sample of ultra-faint dwarf galaxies (purple points in Fig.~\ref{fig10}) appears to follow a different trend to the classical dwarfs. \citet{kaplinghat} have estimated their central densities using the \citet{wolf} mass estimator; the error bars, however, are large because of the small stellar samples for these objects and because a range of NFW parameters is able to reproduce the mass within the half-light radius. If the derived relation for the classical dwarfs also describes the ultra-faints, this would imply that in many ultra-faint dwarfs the \citet{wolf} estimator is biased low by several $\sigma$ and, in some cases, by orders of magnitude. Although this is possible in principle \citep{errani, genina}, we see little justification in modelling the entire population of Milky Way satellites with a unique power law and normalization.

A clear limitation of our comparison to the results of \citet{kaplinghat}, especially for the population of luminous satellites, is the inability of our simulations to match the observed radial distribution of the Milky Way satellites, which is due in part to the fact that host galaxy's baryonic disk is too small \citep{bose_uf}. To that end, comparison of the \citet{kaplinghat} relation to dwarfs selected so as to match the radial distribution of the Milky Way satellites using a procedure similar to that used by \citet{riley} would be more appropriate. Nevertheless, given the various selection biases and the known limitations of both the modelling techniques and the simulations we have employed here, we shall leave a more meaningful comparison of CDM+hydrodynamics simulations to the relation of \citet{kaplinghat} for future work. For now, we point out that an anti-correlation of density and pericentre does exist in our simulations, reflecting ``survivor bias'', but the slope of this relation is weak. Simulations with a more massive baryonic disk are required for a fairer comparison to the Milky Way satellites. If the power-law density-pericentre relation exists and its slope cannot be accounted for by ``survivor bias'', this may pose an interesting challenge for CDM simulations of Milky Way analogues.

\subsection{What is the halo mass of Fornax?}\label{sec:fornaxmass}

\subsubsection{Dynamical estimates}

\citet{readerkal} fit a {\sc coreNFWtides} model to Fornax to estimate its {\it pre-infall} dynamical mass, obtaining $M_{200} = 2.4^{+0.8}_{-0.5}\times 10^9 M_{\odot}$. The pre-infall values of $M_{200}$ for our Fornax analogues lie above this estimate. However, it is challenging to dynamically estimate the halo mass out to Fornax's virial radius ($R_{200} \sim 60-70$\,kpc) from data near its half-light radius ($\sim 1$\,kpc) and the inferred $M_{200}$ depends on the assumed form of the outer dark matter density profile. 

\citet{errani} find that cored, tidally-stripped, models for Fornax favour a higher {\it present-day} total mass than cuspy models (see the left panel of Fig.~\ref{fig9}), with their cored models reaching masses up to $M_{200} = 2.5 \times 10^{10}~M_{\rm \odot}$ within their 68~per~cent confidence interval, while cuspy models reach $2.5\times10^{9}M_{\rm \odot}$. The present-day halo masses of 4 of our Fornax analogues are consistent within 1$\sigma$ with the cuspy models of \citet{errani}. 

\subsubsection{Abundance matching}

An alternative route to estimating Fornax's pre-infall halo mass is from abundance matching. This has traditionally been done using the stellar-mass halo-mass relation ($M_*-M_{200}$), for which Fornax favours a mass $M_{200} \sim 10^{10}\,{\rm M}_\odot$ (\citealt{read_curves}, however, see also the relation of \citealt{miller_abundance}, who predict $M_{\rm200}=3.9^{+2.3}_{-1.4}\times10^9 M_{\rm \odot}$ for a galaxy at $z=0.2-1$ with a stellar mass of Fornax, in agreement with 2 of our analogues). This value is substantially higher than the dynamical estimate of \citealt{readerkal} (though consistent with the estimate of \citet{errani} for Fornax with a large dark matter core). This tension grows worse if we account for the quenching time of Fornax $\sim 2$\,Gyr ago, meaning this dwarf may not have formed all the stars it would have done had it remained isolated. To correct for this, \citet{readerkal} reformulate the halo-galaxy connection as a relation between the mean star formation rate, $\langle {\rm SFR}\rangle$, and halo mass, finding an even higher $M_{200} \sim 2 \times 10^{10}\,{\rm M}_\odot$.

In \citet{readerkal}'s analysis, the SFR is averaged over the time when the galaxy is forming stars. The inferred pre-infall $M_{200}$ for any given galaxy is then dependent on how quenching is defined. The relation is most useful when the mean star formation rate reflects the halo mass of the galaxy (i.e. its ability to accrete and retain gas and to form stars). However, if star formation is enhanced for reasons other than halo build-up, for example, due to a passage
through pericentre \citep{genina}, then $\langle {\rm SFR} \rangle$ could become biased if ${\rm SFR}(t)$ is not truncated appropriately. The recent study of the star formation history of Fornax, using {\it HST} data, by \citet{rusakov} supports a scenario where Fornax has had a recent ($\sim$5~Gyr ago) burst in star formation, which could coincide with its pericentric passage \citep{patel}. In Fig.~A1 of \citet{readerkal}, it is shown that if ${\rm SFR}(t)$ is truncated at around 5~Gyr ago, their
$\langle {\rm SFR}\rangle$-based abundance matching method would give
an $M_{200}$ for Fornax of $\sim 7 \times 10^9$\,M$_\odot$, compatible with the cuspy Fornax analogues we study here.

We also note that, by construction, the abundance matching relation of \citet{readerkal} does not distinguish satellites and isolated galaxies. At the stellar mass of Fornax, the relation is dominated by star-forming galaxies at $z=0$, rather than quenched galaxies. This is in contrast to the population of dwarfs we consider here and could explain the discrepancy between the value of the peak $M_{200}$ inferred via abundance matching and via our hydrodynamic simulations if the halo mass of a Fornax-like isolated dwarf approximately triples between the infall time of a satellite counterpart and $z=0$. However, we have verified that our Fornax analogues in terms of $V_{1/2}$ and $M_V$, which are isolated dwarfs, also have $M_{200, z=0} \lesssim 10^{10}~M_{\odot}$.

Another important aspect of both the $M_*-M_{200}$ and $\langle{\rm SFR}\rangle-M_{200}$ relations is the estimate of the total stellar mass \citep{fornaxdeboer}. Such a quantity is not measurable directly and many photometry-based efforts are susceptible to various selection and modelling biases. A lower stellar mass for Fornax ($M_*<~4.3\times 10^7 M_{\odot}$) would also alleviate the discrepancy between the values of $M_{200}$ predicted in our
simulations and abundance-matching arguments. 

Finally, it is possible that the stellar-halo mass relation in APOSTLE is not representative of the Local Group satellites. The relation is highly sensitive to the subgrid models for star formation and feedback employed in the simulations. APOSTLE uses the EAGLE-REF model \citep{fattahi,sawalapuzzles}, which effectively allows stars to form in lower mass haloes than the EAGLE-RECAL model \citep{eagle,eaglecrain}. Indeed, the latter would place a Fornax-like galaxy in a dark matter halo of z=0 mass $\sim3\times10^{10} M_{\odot}$ \citep{shao}. At such a high halo mass, the tidal stripping scenario for the reduced central density of Fornax would become far less likely (though the scatter in the relation could still permit some analogues to be identified). In this case, different subgrid models (such as those that produce feedback-induced core formation) would be required to explain the low central density in Fornax \citep{navarroeke,readgilmore, pontzen,alejandrocores}.

\subsection{How well is the pericentre of Fornax known?}

The inferred pericentre depends strongly on the model of the Milky Way
potential. This is evident in Fig.~\ref{fig1}, where 7 different
potentials give pericentres between $\sim$25~kpc and
$\sim$150~kpc. These pericentres are derived from detailed models
that take into account the various Milky Way components
\citep{helmiGaia,bajkova}. A pericentre for Fornax that is less than
50~kpc will naturally explain its low density and these pericentres
are, at the moment, not ruled out. Future improvements will come from
accurate mass models of the Milky Way (see \citealt{wang_review} for
an extensive review), which include adiabatic contraction
\citep{callingham2, cautun_mw}, as well as dynamical models that take
into account the potentials of the Large and Small Magellanic Clouds
\citep{patel} and the Milky Way halo growth
\citep{correa1,correa2,correa3}.   

\subsection{Tidal stripping in cosmological simulations}

It has been recently shown by \citet{vandenbosch1} and
\citet{vandenbosch2} (see also \citealt{errani_reconstruct}) that
subhaloes in $N$-body simulations can be subject to artificial
disruption -- that is, disruption due to numerical effects, rather than
physical processes. This could be caused by a gravitational softening
that is too large, as well as discreteness effects, resulting in a
runaway instability. If our simulated subhaloes were to be tidally
stripped prematurely, this could be a potential limitation to our
results.

In Appendix~B we show that, particularly for our Fornax analogues, we
expect the change in the radius of the maximum circular velocity,
$R_{\rm max}$, as a satellites orbits, to be consistent with the ``tidal
tracks'' of \citet{errani_navarro}. We
estimate that the reduction in $V_{\rm max}$ could be overestimated by
$\sim 10$~per~cent. This does not change the fact that our analogues
are similar to Fornax in terms of their present-day $V_{\rm circ}(\rm{947~pc})$ (given that this quantity evolves similarly to $V_{\rm max}$). Moreover, none of our Fornax analogues lose more than $\sim 90$~per~cent of their bound mass and they all have $>10^5$ particles at infall. According to \citet{vandenbosch2} and \citet{errani_navarro}, enhanced tidal stripping due to numerical effects is not noticeable for subhaloes with these properties. Nevertheless, we argue that a comprehensive convergence study of the tidal disruption of haloes in a cosmological setting, with the inclusion of baryonic effects, is certainly warranted.

\section{Understanding the low dark matter density in Fornax}
\label{sec6}

\begin{table*}
\centering
\caption{The effectiveness of various physical mechanisms, within the CDM framework, in explaining the properties of the Fornax dwarf spheroidal, including dynamical properties inferred by \citet{heating} and \citet{readerkal}, and globular cluster (GC) survival. For each scenario we also include the prospect for detecting tidal tails.}
\begin{tabular}{c|c|c|c|c|c|c}
Reason for low density & $\rho$~( < R$_{1/2}$) & Density profile shape & SFR enhanced? & $M_{\rm 200,dyn} = M_{\rm 200, \langle SFR \rangle}$ & Globular cluster survival & Tidal tails? \\ \hline \hline
 \\ \textbf{DM heating alone} & Yes  & Core / Shallow cusp & No  & Yes (for $\sim$kpc cores) & Yes & No \\  \\ \hline 
\\ \textbf{Strong Tides alone} & Yes & Cuspy, disfavoured at $\gtrsim 2\sigma$ & Maybe & (SFR boost required) & Yes & Yes   \\ \\ \hline
\\ \textbf{DM heating + Strong Tides} & Yes & Core / Shallow cusp & Maybe & (SFR boost required) & Yes & Yes   \\ \\ \hline
\\ \textbf{DM heating + Strong Tides} & Yes & Core / Shallow cusp & Yes & Yes & Yes & Yes \\ 
\\ \textbf{+ SFR boost} & && & &\\ \hline

\end{tabular}

\label{table1}
\end{table*}

We now place our findings in the context of the inferred properties of the Fornax dwarf galaxy and summarise the ability of various mechanisms (or their combination) to explain these properties in Table~\ref{table1}.

 Kinematic analyses suggest that Fornax has a low dark matter density, both in its centre and its outskirts, as compared to expectations from abundance matching in $\Lambda$CDM. One explanation for this is the formation of a dark matter core through baryonic processes. Cores can form as a consequence of dark matter `heating' over time due to supernovae feedback associated with extended periods of star formation. For such a model to fit Fornax's low density at all radii on its own, Fornax requires a dark matter core of size $\sim 1$\,kpc and density $\sim 5 \times 10^7$\,M$_\odot$\,kpc$^{-3}$. In this scenario, Fornax inhabits a halo with a high pre-infall mass, of order $M_{200,\rm dyn} \sim 1-2 \times 10^{10}$\,M$_\odot$, consistent with the value inferred from abundance matching $M_{200,\langle \rm SFR \rangle}$ (see Section~\ref{sec:fornaxmass} and \citealt{readerkal}).
 
 Alternatively, Fornax could inhabit a cuspy dark matter halo, with its low density owing entirely to tides. This can occur if Fornax is on a low pericentre orbit around the Milky Way, or if it experienced tidal interactions with other galaxies prior to infall. This scenario requires Fornax's pre-infall $M_{200, \rm dyn}$ to be significantly lower than that inferred from abundance matching with star formation rate. Abundance matching can return a pre-infall $M_{200,\langle \rm SFR \rangle}$ that is biased high if Fornax's stellar mass is overestimated, if its SFR was boosted due to its interaction with the Milky Way, or if the abundance matching relations we have used in this paper are not appropriate for satellites like Fornax (see Section~\ref{sec:fornaxmass}). We note that the analysis of HST data by \citet{rusakov} suggests an SFR boost in Fornax 5~Gyr ago, which may coincide with its orbital pericentre (see the right panel of Fig.~\ref{fig2}). Cuspy profiles are still at $>2\sigma$ tension with the kinematics of Fornax below $R_{1/2}$ (see Fig.~\ref{fig8}), however. A core in these regions, created through dark matter heating by supernovae feedback, for example, would provide a better agreement with the data. 
 
We note that the analysis of \citet{tobeta} has shown that the deviations from a $\rho \propto r^{-1}$ cusp of the inferred density profile of Fornax from \citet{heating}, who used a broken-power-law dark matter density model, is at present consistent with the bias expected in spherical Jeans analysis when modelling an elliptical galaxy (with a Fornax-like on-the-sky ellipticity $e$ = 0.3) inside a cuspy CDM halo that is viewed along its minor or intermediate axis. As demonstrated in Fig.~\ref{fig9}, all of our Fornax analogues are consistent with the axisymmetric Jeans analysis of \citet{hayashi} within the 95~per~cent confidence levels. Analysis using larger spectroscopic samples will clarify whether ellipticity could be the culprit for Fornax's inferred density profile shape. 

In the end, Fornax's low density may owe to some combination of all of the above effects. To determine whether tides have played a role, we can hunt for Fornax's stream of unbound stars. While signs of tidal stripping in Fornax have not yet been detected \citep{walkerfornax,wangfornaxtidal}, we showed in this work that the tidal tails, if present, are expected to be seen over an area of $\sim100$~deg$^2$ with a surface brightness of $\sim35-36$~mag~arcsec$^{-2}$. Such tidal tails should be detectable in future surveys, if tidal effects are responsible -- wholly or in part -- for lowering the dark matter density in Fornax.

Finally, a dark matter core in Fornax has been previously invoked as a solution to the globular cluster survivability problem \citep{goerdt,cole_glob_clust}. In a cuspy halo, the globular clusters are expected to rapidly sink to the centre of the dwarf, while in a halo with a core the globular clusters would stall at approximately the core radius (\citealt{cole_glob_clust}). The survival of globular clusters in Fornax at the present day is thus suggestive of a core in this galaxy. Note, however, that such timing arguments cannot conclusively rule out a cusp \citep{cole_glob_clust,meadows, shao}. In the context of our work here, if tides are important for Fornax after all, we can actually re-ignite the original solution to the `timing problem' for Fornax's globular clusters whereby their infall is delayed purely by tidal effects \citep{Oh2000}.

\section{Summary and Conclusions}
\label{sec7}

Fornax is one of the largest and brightest dwarf satellites of the Milky Way, making it a prime target for spectroscopic and photometric studies that over the years have provided an abundance of data. In particular, measurements of the motions of individual stars in Fornax have been used to place constraints on its underlying gravitational potential. For example, \citet{heating} have found that Fornax must have a low inner dark matter density, which could be a signature of a dark matter `core' in its central regions.

In this work, we have tested an alternative scenario in which the low inferred density in Fornax is due to tidal effects. This was disfavoured by \citet{heating}, who argued that the proper motions, star formation history and inferred pre-infall halo mass of Fornax are inconsistent with this interpretation. We tested this on a large sample of dwarf galaxies selected from the APOSTLE suite of cosmological $N$-body hydrodynamic simulations in Local Group-like environments. We find the following:

\begin{itemize}
\item We confirm that the star formation history of Fornax is
  consistent with that of a dwarf galaxy that is a satellite of the
  Milky Way in our simulations (see right panel of Fig.~\ref{fig2}). \\ 

\item Close matches to Fornax in {\it both} visible magnitude, $M_V$,
  and circular velocity at the half-light radius, $V_{\rm circ}(\rm{947~pc})$, are quite
  rare in our simulations (see left panel of Fig.~\ref{fig2}); however
  we find 9 Fornax analogues (2 of which are particularly close
  matches). These analogues also have inferred orbital apo- and
  pericentres compatible with Fornax data 
  \citep{helmiGaia,fritz,gaia_edr3_pm}. \\

\item By examining mass loss from dwarf galaxies with orbital
  properties consistent with those  measured for Fornax, we show that, even at large present-day pericentres, dwarf satellites can suffer
  significant mass loss. We find that, for galaxies on Fornax-like
  orbits (independently of mass), 26~per~cent lose more than
  50~per~cent of their mass within 1~kpc (see Fig.\ref{fig3}).\\

\item{We explore the relevance of the star formation history to the degree to which a dwarf has experienced tidal effects. We show that in many cases the quenching time
    is directly related to the infall time; however, there exists a
    population of galaxies with similar stellar mass to Fornax, or
    larger, that are quenched long after infall. This is because they are
    able to preserve their gas supply after infall and, in extreme
    cases, even after multiple orbital pericentres (see the left panel of
    Fig.~\ref{fig4}).} \\

\item{We find that in $\sim 37$~per~cent of simulated dwarf satellites with
    orbits similar to Fornax, mass loss begins more than 2~Gyr prior
    to infall. As such, the infall time is not always a reliable
    estimate of the time at which the dwarf had its maximum mass (see
    right panel of Fig.~\ref{fig4}). By examining individual
    Fornax analogues, we found that, in some cases, dwarfs can lose
    mass prior to infall due to interactions with more massive dwarf
    galaxies, perhaps due to infall as part of a group or a
    fly-by. Two consecutive fly-bys are responsible for the tidal stripping of one of our best Fornax analogues (the `star' Fornax on the left side of Fig.~\ref{fig5}).} \\
    
\item{We emphasize that if cores do not form through baryonic feedback in CDM, the low density in Fornax must be due to tidal effects. Although one of our Fornax analogues is a weakly stripped low-mass halo (`thick diamond' analogue), its density profile is discrepant at $3\sigma$ from the density profile inferred for Fornax by \citet{heating} and, unlike Fornax, this galaxy retains $\sim25$~per~cent of its all-time maximum gas content at $z=0$ \citep{spekkens}. If Fornax was stripped to a similar extent as our best analogue, its tidal tails should be visible with a surface brightness limit of $\sim35-36$~mag~arcsec$^2$ and a survey area of $\gtrsim 100$~deg$^{2}$ (see Fig.~\ref{fig6}), which has not been presently achieved with surveys like DES \citep{wangfornaxtidal}. We also show that the proper motions and line-of-sight velocities of the stars in the tidal tails peak at the same values as the bound stars (see Fig.~\ref{fig7}). Similarly to Fornax, our best analogue also exhibits a metallicity gradient \citep{kirbygradient}, leading to the tidal tails which are typically more metal-poor than Milky Way halo stars. These properties could aid the detection of the tidal tails of Fornax in the future. }    \\

\item{By comparing the density profiles of our Fornax analogues to
    that derived for Fornax by \citet{heating}, we show that a NFW profile shape is consistent with Fornax within 3$\sigma$ in the radial range 0.1~-~2~kpc, while Einasto profiles, with shallower inner cusps, are consistent within 2-3$\sigma$ levels. A much better fit to the data is obtained for a stripped dwarf with a rather shallow cusp or core that could have been created by dark matter heating (see Fig.~\ref{fig8}). Our Fornax analogues all have higher values of $\rho_{150}$ than inferred by \citet{heating}, with only two dwarfs consistent with Fornax within 2$\sigma$ and 5 consistent within 3$\sigma$. We note, however, that all of the analogues have values of $\rho_{150}$ consistent with the result for Fornax from \citet{hayashi} within 2$\sigma$. }
  \\

\item{We apply the analysis of \citet{heating}, based on the inner
    densities and pre-infall halo masses of dwarfs, to our Fornax
    analogues. We show that a number of dwarfs with similar $V_{\rm circ}(\rm{947~pc})$ and $M_V$ to
    Fornax, and pericentres between 20 and 120~kpc, have central
    densities comparable to that of Fornax (see
    Fig.~\ref{fig9}). It is clear that there exist dwarf satellites in our simulations, with orbits similar to Fornax, that have been tidally stripped more
    extensively than their pericentres would suggest. We 
    note, however, that a small (< 50~kpc) pericentre for Fornax is not ruled out at present.} \\

\item{Our simulation results are at odds with the inference by 
    \citet{readerkal} of the pre-infall halo mass of Fornax. The
    simulations imply that a galaxy with the present-day properties of
    Fornax should have a pre-infall halo mass of
    $(4-9)\times10^{9} M_{\odot}$, while \citet{readerkal} infer
    $2\times10^{10}M_{\odot}$. One possible explanation for this is that Fornax's star formation was enhanced on infall to the Milky Way. This would raise its $\langle {\rm SFR} \rangle$, leading \citet{readerkal} to overestimate Fornax's pre-infall mass. Alternatively, it may be that APOSTLE provides an inadequate representation of the stellar-to-halo or
    the $\langle {\rm SFR} \rangle$-to-halo mass relation on Fornax
    scales.} \\

\item{We show that the central densities and pericentric distances of satellite dwarfs in our simulations are, at present, compatible with the anti-correlation between these properties in Milky Way satellites highlighted by \citet{kaplinghat}. The anti-correlation in our simulations exists due to the ``survivor bias'', but is relatively weak. The slope of the anti-correlation must be confirmed using Milky Way simulations with a more massive baryonic disk. At present, the limitations of our simulations as well as selection biases and modelling techniques involved in computing the relevant properties for the Milky Way satellites make a direct comparison with \citet{kaplinghat} non-trivial and we shall therefore leave this for future work. } \\

\end{itemize}

In conclusion, we have shown in this work that tides can play an important role in the evolutionary history of Milky Way satellite galaxies, even if that may not be obvious from their present-day orbital pericentres. We have focused on the Fornax dwarf galaxy and have explicitly shown that in CDM simulations in which cores do not form by baryonic processes, the low dark matter density in Fornax can only be explained by Galactic tides. As such, a failure to detect stellar streams associated with Fornax could rule out such models. With regards to the density profile itself, our analysis suggests that the most stringent test of the cusp vs. core dichotomy will come from an accurate inference of the dark matter density profile on $\sim100$\,pc scales. At present, a low-density NFW profile in Fornax
cannot be ruled out, although the data do seem to favour a shallow cusp or core. More precise determinations of the dark matter distribution of Fornax may become possible with more radial velocities in the inner 100-200\,pc, long-baseline proper motion measurements for individual stars with {\it Gaia} and {\it HST} \citep{massari_sculptor,sfw_proper}, accounting for asymmetry \citep{hayashi} and rotation \citep{zhu_sculptor}, and consideration 
of independent constraints from its multiple stellar populations \citep{walkerPenarrubia}.

\section*{Acknowledgements}
 We thank an anonymous referee for a detailed reading of our paper and for several suggestions that helped us improve it. AG and JR would like to thank the organizers of the "5$^{\rm th}$ Gaia Challenge" workshop, where this work originated. This work was supported by the Science and Technology Facilities Council (STFC) consolidated grant ST/T000244/1. AG acknowledges an STFC studentship grant
ST/N50404X/1 and the Ogden Scholarship Fund. 
AF acknowledges support by the STFC [grant number ST/P000541/1] and the Leverhulme Trust. CSF acknowledges support by the European Research
Council (ERC) through Advanced Investigator grant DMIDAS (GA 786910). This work used the DiRAC Data Centric system at Durham University,
operated by the Institute for Computational Cosmology on behalf of the
STFC DiRAC HPC Facility (\url{www.dirac.ac.uk}). This equipment was
funded by BIS National E-infrastructure capital grant ST/K00042X/1,
STFC capital grant ST/H008519/1, and STFC DiRAC Operations grant
ST/K003267/1 and Durham University. DiRAC is part of the National
E-Infrastructure. This work has benefited from the use of {\sc numpy}, {\sc scipy} and {\sc matplotlib}.

\section*{Data availability}
The data analysed in this article can be made available upon reasonable 
request to the corresponding author.




\bibliographystyle{mnras}
\bibliography{fornax} 




\appendix

\section{Spherical approximation to the potential}

\begin{figure}

		\includegraphics[width=.9\columnwidth]{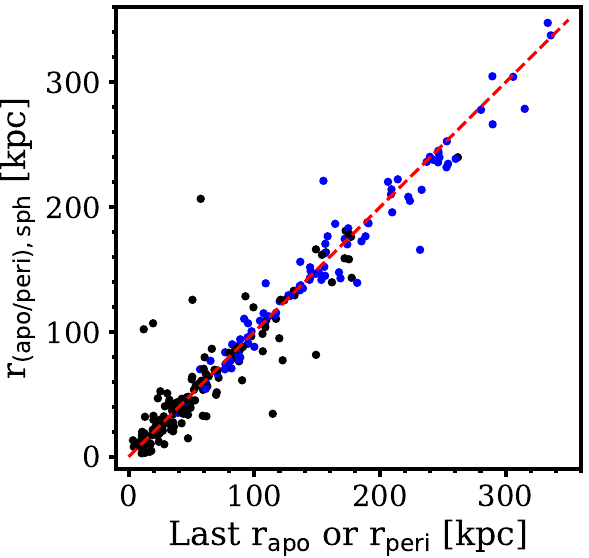}
		\caption{The apo- and peri-centres derived for satellite dwarfs in our sample using a spherical potential approximation, as a function of the corresponding last apo- or peri-centre (computed using cubic-spline orbit interpolation). The dwarfs which recently passed through their apocentre are shown in blue and those that have recently passed through their pericentre are shown in black. The red dashed line is the one-to-one relation. Only the satellites which have passed through their first pericentre are included.}

		\label{figApp1}
	\end{figure}

In this Appendix we show that the spherical approximation to the gravitational potential of APOSTLE haloes is a valid one. From our sample of satellite galaxies, we select those which have already reached their first pericentre. For this sample, we interpolate the orbits using a cubic spline. For the dwarfs which have most recently passed through an apocentre we compare the apocentric distance derived using Equation~\ref{eqperiapo} ($r_{\rm apo,sph}$) with the value obtained using cubic interpolation ($r_{\rm apo}$). For those that have most recently passed through a pericentre, we compare the pericentre values. The comparison is shown in Fig.~\ref{figApp1}, where it is clear that the pericentres (black) and apocentres (blue), derived using the spherical approximation of the potential, follow a one-to-one relation (red dashed line) with the `true' values. There are some noticeable outliers, which correspond to recent galaxy interactions, however, these do not affect the conclusions of this work. Fig.~\ref{figApp1} suggests that the values of the peri and apo-centres derived using the spherical approximation are a good reflection of the true values for these dwarfs. 

\section{The `tidal tracks' of cosmological subhaloes}

In this Appendix we briefly consider how our results could be affected
by numerical artefacts. In particular, we wish to know if 
subhaloes in our simulations are stripped too fast, compared to the
predictions of higher-resolution idealized models in the works of
\citet{pen10, vandenbosch2, dashlibrary2, errani_navarro}. We note,
however, that it is not clear that the results in those papers should
directly translate to cosmological hydrodynamics simulations.

\begin{figure*}

		\includegraphics[width=2\columnwidth]{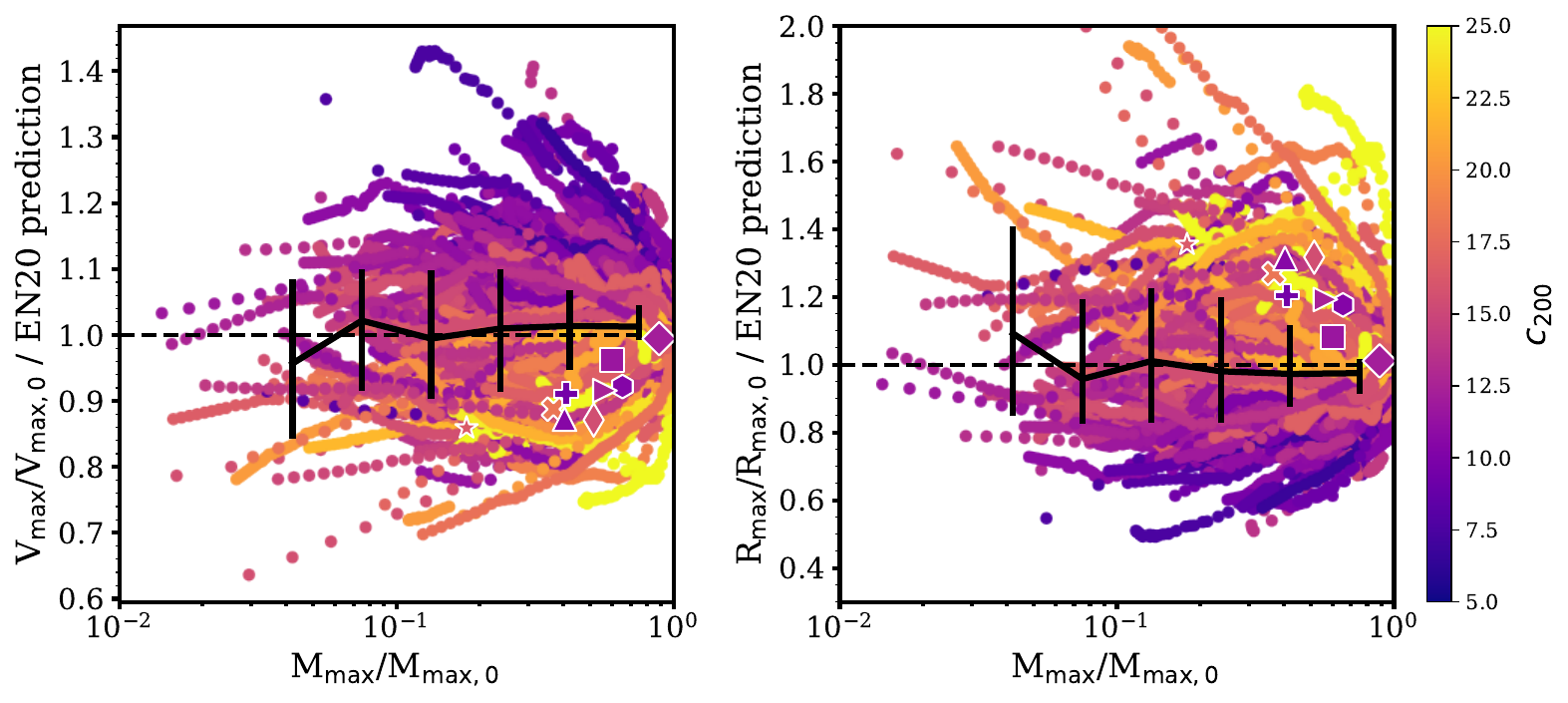}

		\caption{{\it Left:}  fractional change in the
                  maximum circular velocity, $V_{\rm max}/V_{\rm max,0}$, of our
                  satellite dwarfs, divided by the fractional change
                  predicted by \citealt{errani_navarro} (EN20) for a given fraction of bound mass remaining within R$_{\rm max}$,
                  M$_{\rm max}$/M$_{\rm max, 0}$. For each satellite, we use multiple
                  snapshots after infall, but we exclude those where
                  the satellites come within 50~kpc of the host
                  halo. The points are coloured by halo concentration, c$_{200}$. The
                  error bars are the median and 1$\sigma$ intervals in
                  bins of M$_{\rm max}$/M$_{\rm max, 0}$. The various symbols show our Fornax analogues. {\it Right:} as in the left
                  panel, but the fractional change in the radius
                  of the maximum circular velocity, $R_{\rm max}/R_{\rm max,0}$,
                  compared to the predictions of 
                  \citet{errani_navarro}. }

		\label{figApp2}
	\end{figure*}

Motivated by the analysis of \citet{dashlibrary2}, on the left panel of
Fig.~\ref{figApp2}, we show the fractional
change in the maximum circular velocity, $V_{\rm max}/V_{\rm max,0}$, and the radius of the maximum circular velocity, $R_{\rm max}/R_{\rm max,0}$, compared to the predictions of \citet{errani_navarro} for a given fraction of mass lost within $R_{\rm max}$, $M_{\rm max}/M_{\rm max,0}$. The reference values are set at the time of maximum $M_{\rm max}$. Since the $V_{\rm max}$ and $R_{\rm max}$ evolution can be noisy in simulations due to merger and accretion history, interactions with the host galaxy and the particulars of the subhalo finders, we smooth the $V_{\rm max}$ and $R_{\rm max}$ histories of the dwarfs with a 1$^{st}$ order Savitzky-Golay filter over 5 snapshots \citep{savitzky}. The reaction of the circular velocity profile to tidal mass loss in our simulated haloes is generally consistent with the predictions of \citet{errani_navarro}, although the scatter is visibly large and concentration-dependent. Our Fornax analogues, however, clearly deviate from the tracks and from the full sample. Some of this may be due to higher concentration of some of these dwarfs. Additionally, it is clear from the left panel of Fig.~\ref{fig2} that our analogues correspond to a biased selection in $V_{\rm max}$ for a given stellar mass. Nevertheless, considering that the bias in $V_{\rm max}$ is no more than 20~per~cent,
our Fornax analogues are still consistent with the value of $V_{\rm circ}\rm{(947~pc})$
of Fornax (assuming that this changes similarly to $V_{\rm max}$).

It should be noted that the above is only a conservative estimate of the effects of artificial disruption on our simulated Fornax analogues. An interesting feature of Fig.~\ref{figApp2} is that the reduction in
$R_{\rm max}$ and $V_{\rm max}$ is clearly dependent on halo concentration,
$c_{200}$ (identified by the colour of the symbols in Fig.~\ref{figApp2}). This is consistent with the results of
\citet{dashlibrary2}. Moreover, \citet{errani_navarro} point out that the behaviour of the `tidal track' is orbit-dependent, so we do expect some scatter about their median relation. We also note that our Fornax analogues satisfy the force softening and number of particles criteria of \citet{vandenbosch2}, which these authors suggest leads to an evolution of bound mass accurate to 0.1~dex. Furthermore, it has been shown that tidal tracks of aspherical haloes deviate from the spherical cases explored in \citealt{vandenbosch2} and \citealt{errani_navarro} \citep{sanderstracks}.

\bsp	
\label{lastpage}
\end{document}